\documentclass[10pt]{article}
\usepackage{amsfonts}
\usepackage{graphicx}
\usepackage{epsfig}
\usepackage{color}

\begin{document}

\title{Thou shalt not take sides:\\ Cognition, Logic and the need for changing how we believe}
\author{Andr\'e C. R. Martins \\
	NISC -- EACH -- Universidade de S\~ao Paulo,\\
	Rua Arlindo B\'etio, 1000, 03828--000,  S\~ao Paulo, Brazil}

\date{amartins@usp.br}

\maketitle

\begin{abstract}
	
We believe, in the sense of supporting ideas and considering them correct while dismissing doubts about them. We take sides about ideas and theories as if that was the right thing to do. And yet, from a rational point of view, this type of support and belief is not justifiable at all. The best we can hope when describing the real world, as far as we know, is to have probabilistic knowledge, to have probabilities associated to each statement. And even that can be very hard to achieve in a reliable way. Far worse, when we defend ideas and believe them as if they were true, Cognitive Psychology experiments show that we stop being able to analyze the question we believe at with competence. In this paper, I gather the evidence we have about taking sides and present the obvious but unseen conclusion that these facts combined mean that we should actually never believe in anything about the real world, except in a probabilistic way. We must actually never take sides because taking sides destroy out abilities to seek for the most correct description of the world. That means we need to start reformulating the way we debate ideas, from our teaching to our political debates, if we actually want to be able to arrive at the best solutions as suggested by whatever evidence we might have. I will show that this has deep consequences on a number of problems, ranging from the emergence of extremism to the reliability of whole scientific fields. Inductive reasoning requires that we allow every idea to make predictions so that we may rank which ideas are better and that has important consequences in scientific practice. The crisis around $p$-values is also discussed and much better understood under the light of this paper results. Finally, I will debate possible ideas to try to minimize the problem. 
		
\end{abstract}

\section{Introduction}

We believe. We believe in the honesty of certain people and in the dishonesty of others. We may believe we are loved. Some of us believe in deities, some believe in political or economical ideas, some believe in scientific results. We have beliefs on how the world ought to be and, at least for those beliefs, we are often very aware that reality does not correspond to them. But we also hold beliefs about the actual state of the world. And sometimes these beliefs are so strong that we feel justified in saying that we know. Even when they are not that strong, we often choose to say and defend those beliefs as if they were indeed true. We can also hold beliefs about abstract entities that do not actually exist in the world outside our minds and languages, such as numbers or deities. We make claims of knowledge, and we expect to have real knowledge, to the point that in Logic, it is often considered that the phrase ``Someone knows $p$'' means that $p$ is actually true \cite{faginetal95a}.
The beliefs we defend have a vast importance in all social aspects of human life. We have several organizations based on the spread of ideas, on defending the beliefs that the people in those organizations hold to be true, as if they were superior to everyone else's beliefs. We join groups that share our beliefs and we fight (sometimes only verbally) those we disagree with. But, as we will see, taking sides is not only logically wrong, it also makes us dumb, as experiments have been showing lately.

The actual question of what beliefs and knowledge are is and old and, maybe surprisingly, a still open problem in Philosophy \cite{shope02a}. The concept of belief seems so natural that we have never really doubted its usefulness and we only tried to understand and define it properly. But, when we have asked what knowing and believing are, we have usually been forced to ask if knowledge is even possible or if Skepticism \cite{klein14a} is actually unavoidable. And yet, whenever we have taken a position that did not included the hardest forms of Skepticism, the concept that we would accept some ideas and reject others have come naturally. We do not question it. This concept, accepting and rejecting ideas about the real world, seems very natural and, indeed, it has survived for a long time without being properly challenged. It could be the case we have never challenged it just because it is too obviously right. But it is a scientist duty to ask if this is indeed the case. And, when asking that question, we must acknowledge that the concept might just feel obvious as consequence of some very deep instincts we have. 

Indeed, recent results and experiments in Cognitive Psychology point to serious problems with the way we naturally work with our beliefs and how we update them.  Our many flaws in reasoning have led to the conclusion that the main function of our reasoning might not to make better decisions. Our reasoning about ideological issues does seem to come in packages better described as irrationally consistent \cite{jervis76a}. Instead of looking for truth and correctness, it seems clear now we use it to win arguments \cite{mercier11a,merciersperber11a} and to protect our own cultural identity\cite{kahanetal11}. Looking for the correct or best answer might even be irrelevant in too many cases and, at best, a possible use for our reasoning skills that is nothing more than a side effect of the motives that have led evolution  to give us something we call reason and the ability to argument. Our reasoning skills seem to have social purpose as their most important, central function. Their main function seems to be to allow us to identify and defend the ideas of those groups that define the way we see ourselves.

These results pose interesting questions about the way we work and defend our most fundamental beliefs, the same beliefs that we feel define who we are. Since we do not use our reasoning to find the truth under these circumstances, a natural question we should ask but usually avoid is how we could counter this insidious bias. But even this question might not address the problem fully. A more meaningful and complete question could actually challenge the very idea that we should have identity-defining beliefs. That is the point of view I will defend in this paper. Evidence shows clearly that when we embrace an idea, be it a religion or a political ideology, our brains start working to defend our interest and that what we are interested is to advance the cause of the ideas and sides we  chose, instead of looking for the correct answers. While it is not clear if this could pose a problem for moral issues, where no correct answer is known to exist, we do have a problem when confronted with any type of ideologies that has claims on how the world actually works. Here, I will show that why we should make very strong efforts to avoid those types of belief, that we really should not take sides. In Section \ref{sec:why}, I will present the problem as we know it now. I will review some of the evidence that shows our many reasoning errors and biases and proceed to investigate, from a normative point of view, if there is any reason to hold and defend beliefs. 

In Section \ref{sec:extreme}, I will show that our taking sides also have very important effects of the appearance and maintaining of extremism. I will investigate, by using the tools of Opinion Dynamics  \cite{castellanoetal07,galametal82,galammoscovici91,sznajd00,deffuantetal00,martins08a} how the way we treat beliefs can have a profoundly deep impact on the extremism of our own thoughts. And how a subtle change in the way we think about problems, in particular if we are look for absolute or proportional answers, can have an absurdly strong effect on extremism. To explore this problem, I will investigate the effects of two possible causes: the differences that arise from communication methods where uncertainties are expressed, and the differences on how strong opinions can become based on how our own internal expectations about the information we obtain. As we will see, the simulations presented here will make a very strong case in favor of why we must change the way we deal with beliefs, not only to become better when looking for true or best answers (as if was not already reason enough) but also to avoid extremism. The simulations will show that, regardless of the way information is presented, what actually drives extremism is the expectation to find a definitive single answer instead of being satisfied with a mixed result where two competing ideas can add something to final solution.

In Section \ref{sec:science}, I will address the consequences for scientific practice in general that we can learn from the conclusion that we must not take sides and must not hold identity-defining beliefs about the real world mean. As we will see, avoiding taking sides can help correct some problems we still have with the reliability of the scientific enterprise. Basically, the question of whether or not scientific knowledge can be considered more reliable (or, better yet, far more reliable) than any other kind of knowledge is one that many people consider settled. And yet many of those people who think they know the answer strongly disagree with each other. They take sides and defend those sides. Individual points of view vary as widely as from considering that all scientific knowledge is no more valid than any other type of opinion, as in some wild interpretations that followed the proposals in the Strong Program in the Sociology of Knowledge\cite{bloor91a}, to very strong realistic beliefs that we are actually uncovering the true laws of Universe. Or even bolder positions that allow Stephen Hawking to claim, maybe as a  joke, that he believes Roger Penrose to be a Platonist\cite{hawkingpenrose10a}. While it would make no sense to deny that actual scientists are human beings and, therefore, subjects to our failures and all types of social pressure, it also makes no sense to deny the incredible achievements we have and attribute them to sheer luck. Still, we will have already learned that we should be quite wary of ideas so strongly defended by their proponents as those of both sides of the debate do. It might indeed be wiser to learn which aspects of the best description of the problem each of the sides got right. In the terminology of  Section \ref{sec:extreme}, we should strive here to be mixers instead of wishers.

	That we have sent probes to the outer reaches of the Solar System, that we have eliminated diseases and dangers and we live much longer than we used to just a couple of centuries ago are clear indications that we must be doing something right. On the other hand, if we consider the notion that, from a rational point of view, we should not take sides and that, when we do, those beliefs can seriously compromise our ability to reason, we do have a solid cause to be concerned about scientific work (or any work that requires intellect). Indeed, instead of a cold and strictly rational relationship with the word of ideas, scientists are actually quite passionate about what their defend. Worse yet, we now have access to data that points to the fact that stronger intellects might actually show stronger polarization of opinion effects\cite{kahanetal12a}. This suggests that more intelligent people use their intelligence not for finding the truth or even simply to honestly test ideas. Instead, once more, it seems we all use our brain powers to protect the things we believe in and, the larger the brain power, the better defense one can prepare. But, maybe a little unexpectedly, this effect can be even stronger, the better prepared an individual is. At one side, we have our achievements to tell we are doing something right, at the other side, evidence that consensus might just be a social construct.

	An analysis of the question on what Science can prove from a logical point of view does not get us more confident, if we only use a deductive point of view. Popper idea of falseability \cite{popper59} was indeed based on the trivial logical fact that when a theory makes a prediction, just observing  the prediction holds provides no proof at all that the theory is right. On the other hand, if the prediction fails, we do know that, since the conclusions were wrong, something must be wrong with the premises. For Popper, that meant that, after proper checks for other possibilities, the theory had to be false. But that is actually only true if too many other conditions are met. First, there must be absolutely no chance for experimental error. And it must also be the case that you consider as theory the whole set of ideas you used to make the prediction, including every detail about the Universe you assumed during your calculations. If you use a more traditional meaning to the word theory, Popper idea that a theory can be proved false is just wrong from a strict deductive point of view \cite{quine53a,duhem89a}. The discovery of Neptune is the classical example of a prediction (the movement of Uranus) that was wrong not because the theory of Newtonian Gravity was failing. Uranus did not behave the way the theory predicted because there was another unknown planet we did not know at the time, Neptune. Neptune gravitational influence could not have been considered in the calculations before its existence was known. Auxiliary hypothesis like the structure of the Solar System are not considered part of Newtonian mechanics or any theory of Gravity. The actual failure of any predictions can be due to the theory we are testing being wrong, of course. But the failure can also be caused by many other hypothesis we had to make to obtain the predictions. That is, from a strictly deductive point of view, we can not hope to prove any theories neither right nor wrong. The existence of Neptune was verifiable and, therefore, it might seem at first that we could actually have disproved Newtonian Mechanics if Neptune did not exist \cite{laudan90a}. But other hypothesis could have been made then. And, at the very least, any real calculation always involves approximations in the description of the real world. As such, we never really have what would have been an exact prediction from the theory. There will always be the possibility that a new, unthought hypothesis could save any given theory.

	The real problem is, of course, as we will see in Section \ref{sec:why}, the fact that Deduction can not actually tells us anything about the real world. This is true for scientific reasoning, just as it is true for any other kind of reasoning. Instead, we need inductive tools to estimate which theory is more likely to describe the world correctly. Indeed, more current descriptions of scientific work have proposed that scientist would behave as if they were Bayesian agents, an idea known as Confirmation Theory \cite{earman92,maher93,jeffrey04}. While this is a solid prescription for how scientists should behave, there is actually little evidence that scientists reason better than any other intelligent person. While a Bayesian description does capture the qualitative aspects of our reasoning nicely \cite{martins06,tenenbaumetal07}, we know we actually fail at most probability estimates\cite{plous93,baron,kahneman11a}. Indeed, since the first experiments on how we humans change opinions when presented new data, it has been observed that we have a strong tendency to keep our old opinions, even when we already are talking about probabilistic opinions, a bias known as conservatism \cite{philipsedwards66}. It is quite likely that scientists would exactly the same, even more when we consider that specialists in an area are better equipped to actually defend their points of view, despite the fact that they should actually test those points of view, instead of defending them. This means that it is reasonable to expect that we will have problems in Science, problems caused by the human failures and biases described in this paper. Therefore, we must ask whether these problems might be minimized by some of the characteristics of the scientific work, by social effects, or not at all. If they are, understanding which characteristics are important for that minimization of the problem could, in principle, help us improve the way we reason, as scientists, as citizens, and in our general lives. If it is not, we must learn to improve and correct the ways we, scientists, work.

	We already know that previous observations of scientific work have suggested that scientists break some expected norms all the time. Surprisingly, we will see that part of this non-compliance is actually in complete accordance to normative Bayesian reasoning and it might actually be one reason why some areas are quite successful in explaining their objects of study. I will show that, while those observations, when they were first made, were considered evidence that we should adopt a vision where all ideas are equally valid, this proposed (and quite damaging) Relativism is not justified at all. The concept that we should accept the creation of new ideas, regardless of how much they contradict what we currently know is actually a good prescription. But Induction dictates that all these ideas must be ranked in a probabilistic 
	way and that means they will not be equal at all. That quite obviously means that some ideas will be far more probable than others. In particular, there will be theories that we can claim, in a harmless abuse of language, that we have proved to be wrong. The abuse comes from the fact that there will always remains a technical, ridiculously small probability that they might actually be true.	
	On the other hand, ideally, we have to remember that we should keep even those theories as very remote possibilities. It is always justified to discard them in cases where our limitations dictate we can not consider all possibilities and must focus only on the more probable ideas, of course. But, from a scientific and epistemological point of view, no idea can actually be discarded by data and they should survive in a limbo of basically useless concepts. And that means, among other things, that we actually need pure theorists in every mature area and those theorists should be capable of providing testable predictions.
	
	Finally, not taking sides has also another very important consequence on how we should analyze data. We have created and we have promoted the widespread use of statistical tools that have the objective of discarding hypotheses. Far worse, we have made those tools central to the whole process of acceptance of ideas in some areas of knowledge. This human desire to know the absolute truth and avoid dealing with the undesired uncertainty of the probabilities is at the very core of the whole problem caused by the widespread use of $p$-values and tests of hypothesis that now we know are making whole fields of knowledge far less trustworthy than we can tolerate. We must learn fast that we cause too much harm by using techniques aimed at excluding ideas, instead of ranking them. And we must change as fast as possible what is considered proper statistical techniques since many of those techniques only make our biases and desire to defend an idea a much stronger problem.

\section{Why we should not hold any beliefs at all}\label{sec:why}

\subsection{Individual and Group Reasoning}

That we humans fail at reasoning far more often than we would like to believe is now a well established result. The so called paradoxes of choice of Allais \cite{allais} and Ellsberg \cite{ellsberg61} have shown that our decision making is not rational, in the sense that it can not be described by a simple maximization of any utility function. We actually behave as if any probability value told to us were subject to the application of a weighting function before we use the value to make any decisions \cite{kahnemantversky79,birnbaum2008a}. Among other things, that means we overestimate the probability of unlike events. We see correlations in data when there is none simply because they seem to make sense \cite{chapman69} and fail to see them when they are surprising \cite{hamiltonrose80}. We interpret probabilistic information too badly, even in professional settings. To the point that we already have efforts to improve the understanding of basic probabilistic concepts among health professionals \cite{gigerenzermuir11a}.

Since the first experiments with probabilistic choices, scientists have also explored other aspects of our reasoning with results that can be considered even more disturbing. We actually fail at very trivial logical problems \cite{watsonjohnsonlaird72a,tversykahneman83a} and we provide different answers to the same question simply because it was framed differently \cite{tverskykahneman81}. We are not even aware of how badly we reason, we overestimate our chances to get correct answers \cite{oskamp65a}, with the possible exception of the cases when we are actually really good at those specific questions, when underestimation of successes can happen \cite{lichtensteinfischhoff77}. These two effects combined can generate a dangerous combination of overconfidence among the incompetent and underconfidence among competent people and the consequences of this could be potentially disastrous. A curse of incompetence seems to plague us in some areas, and this can be particularly hard to notice for the cases where the competence to perform a task and the competence to know if we are competent are exactly the same ones. That basically makes it impossible for those who are not trained to actually know that they should know better, since they lack the competence to judge their own competence \cite{dunningetal03a}. And there is more. Sometimes, as for example when presented with new information about a problem, we can become more certain of our answers even when we are making actual worse evaluations \cite{halletal07a}. When reasoning in groups, we, at least, have mixed, good and bad, results. Our combined reasoning can indeed be more solid under some circumstances, a phenomenon known as the wisdom of the crowds \cite{surowiecki05a}. But interaction and social pressures inside groups can actually make groups reason far more poorly than individuals \cite{janis72a,kerretal96a}. Social pressure can make us fail even in absurdly easy tasks almost nobody would get wrong, as shown, already in 1955, by the Ash experiments \cite{asch55a,asch56a}.

The combination of all these experiments paints a very grim scenario about our reasoning skills. When compared to all our accomplishments as a species, it should be clear that a part of the answer is missing. More specifically, how and why we can be so bad at reasoning as we actually are and still have advanced to heights no other species could either dream of. Of course, there is another side to this story, important facts that makes us sound less utterly incompetent. In particular, there are fundamental issues associated with the fact that information processing is a costly effort.  More brain power requires more energy consumption. That means that theoretically optimal decision rules might not be ideal in the real world, where a compromise between effort and correctness might be the actual ideal \cite{simon56}. So, if there are situations when we can get a reasonably good answer most of the time with less effort, by using some simplified rule of thumb, that informationally cheap answer might be the optimal solution in the evolutive problem, even if it is not the most accurate answer to the problem. By spending less energy thinking, it might have been in our best interest to do not explore the problem further when a decent answer is available \cite{gigerenzertodd00a}. Indeed, there are quite simple heuristics that have been shown to be very effective \cite{tverskykahneman73a,gigerenzergoldstein96} at giving good, reliable answers. Our brains might just be looking for workable approximations, not the correct ones.
Indeed, our reasoning seems to be clearly not well adjusted to logical problems at all, not without some amount of real previous training. But, while we are naturally incompetent at solving novel logical problems when formally identical problems using situations we are familiar with are presented, we 
are actually quite good at solving them \cite{johnsonlairdetal72a}. Even our probabilistic biases can actually be understood as reasoning that is similar to a Bayesian inference problem \cite{tenenbaumetal07} where we correct the probability values we hear assuming they might be just an estimate \cite{martins06}. It is in new problems, those far from our daily life, that we seem to fail even in the simplest questions. The worrisome thing to notice here is that all scientific advance happens far from problems we are familiar with the answers.

More recently, a very interesting answer to why we  actually reason and, therefore, why we seem to have so many biases and do so many mistakes,  have started to become clear. 
While reasoning can actually be used for looking for correct answers, this might not be its primary function. The ability to look for correct answers might even be just a luck accidental consequence of our actual need and use for reasoning. The more we learn, the more it seems we use reasoning mostly as a tool for winning arguments \cite{mercier11a,merciersperber11a}. When we have one central belief we defend, one idea we want to claim right, we tend to believe in everything that would support that conclusion and deny everything that would go against it, even when those beliefs are logically independent \cite{jervis76a}. The central point here is that winning an argument does not necessarily means being correct. Not at all, it just means we want to convince the listeners or be convinced by them. When there is already agreement, there would be no need for further exploration of the problem. That explains the very well known confirmation bias \cite{nickerson98a} where people tend to look solely for information that confirms what they already think. Interestingly, this tendency to use reason mainly as a social tool to convince others is a result that appears to be universal, and not dependent on specific cultures \cite{mercier11b}. That is, the real function of our reasoning abilities would be the social one. Our quest for truth, the function we would like to believe, might have nothing to do with our reasoning.
Of course, winning arguments is not something we all can do. There has to be something a little more to it and we can guess what it could be if we realize that when someone inside a group wins an argument, whomever that person is, the whole group will tend to shift to a common opinion everyone will hold. That opinion can eventually be associated with that group, even to to the point of being a way to identify the people who belong to that group.

We must also realize that our reasoning flaws are not restricted to our everyday problems, with no influence on expert opinions or how they are perceived. We already have evidence that the public perception of scientists opinions can be quite wrong exist in several issues, such as global warming\cite{goebbertetal12a} or the disposal of nuclear waste \cite{slovicetal91a}. Kahan et al have observed that even the way we perceive whether there is a scientific consensus on a specific question is influenced by which behaviors we find socially acceptable and which behaviors we believe are socially detrimental\cite{kahanetal11}. We know we value consistency in our beliefs to the point that the set of beliefs we adhere to can be more easily described as irrationally consistent \cite{jervis76a}, since we accept independent ideas as a package, put together for one sole purpose, that they support a conclusion we wish to be true. We call these packages ideologies and we take sides and defend them. We often define who we are based on the ideologies we prefer. But, by doing so, we condemn ourselves, even the brightest ones among ourselves, to irrationality and stupidity. Indeed, there is now good evidence that our reasoning works in an identity-protective way\cite{kahan13a}. Given the amount of already existent evidence on how our sides influence our reasoning, Kahan has proposed that information should always be presented in ways that are compatible with the values and positions of the listener, in order to allow that new information to be absorbed and analyzed with less prejudice \cite{kahan10a}. The cleverer we are, the better we can create arguments to defend our identities and the harder it may be to learn we are actually wrong! That suggests once more that, when facing problems that are associated with the ways we define ourselves, the function of our reasoning is not to find correct answers. Instead, its real function is to protect that identity, to protect our group and our side of the discussion. That includes, as we have discussed, forming arguments that support our views. Indeed, tests of how scientific literacy as well as numeracy correlate with beliefs about climate change among member of the public showed that those who scored higher on literacy and numeracy presented the strongest cases of cultural polarization \cite{kahanetal12a}.  The increased skills actually allow people to defend their chosen positions better, instead of making them better at analyzing the literature and the evidence. The conclusion we can arrive from those observations is a scary one. Brilliant individuals who take sides might not be trustworthy about their opinions at all. Their arguments might be worth listening to and analyzing. But their final conclusions must be seen as irrelevant, at least when they show certainty.

\subsection{Normative Reasoning}

Certainty (or the desire for certainty) is indeed a central part of the problem.
We, as scientists, should want our reasoning to help us to arrive at the right answers, we assume that is what reasoning does, instead of simply being a tool to advance the goals of our own specific social group. That is something we already believe to be fundamental. Since commitment to one idea seems to make all of us reason poorly, it makes sense to ask if we should not just get rid of such commitments. To move forward in answering this question, we must now ask if, from a normative, logical point of view, such commitments even make sense. It is true that  we are so used to choosing ideas and defending them that we do not even question if we should keep doing it. We do not ask if there are better alternatives, we have never questioned if the concept of taking sides might not just be plainly wrong. But that is exactly what we need to do.

Defending an idea would make sense if we knew it to be THE truth. So, the first thing to consider is if, when we claim to believe in something, we are actually saying we have concluded that something is true.  That we know it to be true, correct, and not either an error nor a deception. There is, of course, a weaker meaning to the word believe, as when we say ``I believe it will rain tomorrow''. In this case, we are not really claiming knowledge, we are just making a statement that can be loosely described as probabilistic, despite the fact that no probability was mentioned. Still, there is clear uncertainty in the phrase. As such, this is not the meaning we are interested about, since our social group will not suffer and our identity will not be threatened at all if it turns out we were wrong. We do not really take sides on those questions and we easily accept those beliefs can be wrong. It is when believing is used as if it meant knowing that we can be in trouble. Therefore, that is the meaning we must investigate.

The problem of what knowing something means is actually a very old philosophical question. And, despite several attempts at defining the meaning of knowing something, no clear answer has ever been reached. Apparently reasonable answers, such as the concept of justified true belief, are known to have flaws and exception cases that make them not a real correct answer \cite{gettier63}. Deductive Logic exists at least for as long as since Aristotle, allowing us to make proofs where no doubt exist, but only as long as we do not doubt the premises. Newer, more complete, and better versions of Deductive Logic have appeared but they all depend on what you accept initially as true. There is always the need to already believe some initial concepts as true in order to use the logical tools, just as we need postulates to start any kind of mathematical thought. First truths seem to be unreachable by any Deductive Logic, as far as we know today. We can prove conclusions, we can not prove premises, not unless we make more primitive premises. So, we must ask how can we ever accept those initial premises, how we can ever learn something about the real world.

The form of reasoning we have that allows us to accept propositions about the real world is Induction. As far as we know, it is as old as Deduction, but it has never led to certainties. Indeed, it is based, among other things, on the assumption that the patterns we have observed in the past will be repeated in the future. And that assumption will not be true in every situation \cite{hume,goodman46}. While it is reasonable to expect Gravity to keep working as it has always done (we do not know it will, Induction offers no proof of it, but that conclusion is still very reasonable), there are actually areas of knowledge where expecting the future to strongly resemble the past sounds as an absurd, such as technology, the behavior of financial markets where completely new products are always been created, or the evolution of our own societies and economies as more and more types of work can be replaced by machines. Induction can still be useful in these cases, as we try to understand some underlying laws that might be more basic than the superficial data but doubts always remain.

Indeed, from what we have learned so far about its uses, inductive reasoning can be a powerful and very useful tool, but it means that the desire for knowing the truth must be abandoned. In its place, a probabilistic approach, where we only estimate the likelihood that a given statement is true, is required. How to estimate those probabilities and change them as we learn more can be a very difficult task in real problems, but, at least in theory, we have the prescription for how it must be done. More than that, it is interesting to notice that different assumptions on the same problem of how induction should be performed lead to the same rule for changing probabilities as we learn more about the world\cite{ramsey31a,jaynes03,caticha04a}. We just need to apply a deceptively simple probability rule known as the Bayes Theorem. However, despite Bayes Theorem apparent simplicity, using it in the context of real problems is almost never easy. Far from it, we still are plagued with a large number of unsolved problems, some of them related to the specification of initial knowledge, others about the models about the world that are required to use the theorem for an actual complete description of the possible theories about the world.

An  approach that is equivalent to the Bayesian method was proposed fifty years ago by Solomonoff \cite{solomonoff64a}. This approach sheds some very interesting light on the full requirements of the Bayesian Induction. For a full analysis of any problem, our probabilistic Induction would require both infinite information processing abilities to generate all possible explanations that are compatible with the data we have, as well infinite memory to deal with all those explanations and use them to calculate its predictions. While those requirements do not make Bayesian methods wrong nor useless, they do mean that even when estimating probabilities, the best we can hope for is an approximation to the actual probability values. Uncertainty remains even about  the ``real'' probability values and better methods to estimate uncertainty can always be created to get us a little closer to the best possible answer.

The state of the area and the technical difficulties of actually using those methods are interesting problems, but beyond the scope of this paper. What matters here is that we have learned that, given our current state of knowledge in Deductive and Inductive Logics, probabilistic beliefs about the world actually make normative sense but certainties do not. That is exactly the way Bayesian methods work. While we sometimes use the word belief in a probabilistic sense that fits very well with inductive considerations (I believe I will get that job), there are also times when we use the word belief as a weird claim of unjustified certainty. There are times when we decide to take sides about things we can not be certain about. When we say someone believes in some set of religious claims or in an ideology, for example, we are often saying that person consider those sets of claims to actually be true. Some people would agree that certainty is not really warranted but that the set of beliefs is that individual best guess, a guess that person is willing to defend. And that willingness to defend an idea is far stronger than claiming we just think the idea to be very probable. It is one thing to behave like that in moral issues. It is a completely different problem to do the same on descriptions of the world. The world is the way it is, regardless of what we think or feel about it. To make strong statements about the world, when those statements are simply not warranted should be considered a serious form of irrationality. And a damaging one, since our own psychological characteristics will make those choices turn us into far less intelligent versions of ourselves. 

Things get even weirder when we notice that most sets of beliefs are actually logically inconsistent, with huge holes that their defenders prefer to ignore. Still, even if they were not inconsistent, we would still have problems, as we do not have the tools to say we know the truth on the claims they make about the real world. There is nothing in our current logical or philosophical knowledge that says that certainties such as those are even remotely possible. What we do know is that belonging to a group, any group, can make sense as a social choice and that this belonging affects our intellect. Accepting the group premises as true, whatever those premises are, on the other hand, is actually an irrational choice. We are probably far more efficient truth-seekers if we never take sides in this kind of irrational battles.

\subsection{Psychology experiments and Logic combined}

It is worth to review the argument, and put it all together now. Actively believing and defending an idea brings a series of negative consequences to our ability to pursue the truth. While defending our points of view, experiments show that we accept concepts simply because they support the conclusions we wish were true. At the same time, we submit the ideas that we disagree to an actual rational analysis, looking for possible errors. We value the ideas we believe in ways that resemble an endowment effect \cite{thaler80a}, where we value the same object much more if it is our possession. While the experiments conducted so far were focused mostly on political views and popular opinions about scientific issues, their results seem consistent enough. That consistency makes it very reasonable to expect the same problem should be observed in any ideas we claim to believe. Indeed, that is exactly the way the world seems to operate.

The special case of moral beliefs could be treated apart, as it might, at least in theory, be possible that a person actually knows her moral preferences. But we have a serious problem when we consider beliefs about the real world. The choice of one idea over another is never justified from a rational point of view, at least not to the point of certainty. In situations of limited resources, choosing one idea because we have no time to consider all other alternatives is a perfectly rational choice. For example, choosing one treatment for a patient is necessary, but that does not mean that we decide all other possible illnesses must be wrong. We just decide the best current course of action, based on the chances we estimate for each disease and how serious their consequences might be. That means that the correct thing to do is not taking sides and defend one specific diagnosis, the decisions we make today in this case must be subject to future change, as we learn more. It is very clear that any physician who would choose one initial diagnosis and then simply defend that choice and never change it should be considered incompetent, damaging and dealt with accordingly. Things are not different for any other area of knowledge. Decision making is fundamental and unavoidable. But we do have the tools to deal with decision problems involving uncertainty.

Of course, this whole analysis assumes the debate is happening between knowledgeable, honest people. When debating people who refuses to even learn about the evidence that already exists, taking sides might be almost unavoidable. But that should not mean defending one idea as true, even though things might look that way. It can be just a very basic defense of the need for rational thinking as opposed to the very unreliable human thinking. In such cases, the decision process might eveb tell us the best thing to do can be to take the side of competence and clear logical thinking against cherry-picking tactics and circular reasoning. But it can also be true that the best tactics might be simply to ignore the irrational debaters. The discussion in these cases actually needs to be far more fundamental, if it will happen at all. It must actually be about reasoning in minimally competent ways, about avoiding human biases, and also looking at the real complete data before making a probabilistic estimate. Rationality requires us to consider every possible idea as a possibility, but some ideas will be incredibly improbable. In any circumstances, however, if one group of people simply choses one specific idea they will support and defend, regardless of evidence and rational argumentation, the only possible next step might be to label those individuals as irrational and ignore them, if possible. As they are not looking for the truth, they are simply using our human very bad natural reasoning to defend their group.

In all the necessary considerations, we should, of course, not forget that our brains are limited machines. Some ideas are so improbable that we can claim them wrong in any practical terms based on the evidence we have collected so far (we will see an example in Section \ref{sec:science}). In those cases, it will often make sense to not consider concepts that we have sufficiently strong evidence to discard as highly improbable when making decisions. But this is different from claiming our ideas to be correct. Or even to claim certainty that the ideas we have discarded are wrong, since it is always possible that new evidence might force us to change our opinions in the future. We, scientists, claim to actually know that and that fact that ideas might return and that we always should be open to changing our minds when faced with that new evidence. However, experiments show that by belonging to any group where an idea is defended we make ourselves vulnerable to our human shortcomings. Observations of real scientists working actually agree that we are not different from the picture that emerges from the experiments. We are far too stubborn and we cling to the ideas we defend despite evidence, as we will review in Section \ref{sec:science}. We are not different and, as any humans, we stop being truth-chasers and become idea-defenders. That actually makes our individual opinions unreliable when we take sides. Anyone that is out there defending an idea should actually be considered a non-reliable source of information. That is why we must start as soon as possible to guard ourselves against taking sides.

\section{How the ways we handle beliefs and how we communicate can influence extremism}\label{sec:extreme}

Unfortunately, our desire to have one side to defend does not only makes us unreliable. It can also be a fundamental factor in the appearance and spread of extremist views.
Extreme opinions can be observed in many different issues and they are often not a problem. While a very strong opinion is sometimes warranted (for example, about not leaving the $20^{th}$ floor by the window), there are several other circumstances where the same strength of opinion makes no sense, such as many problems where no scientific consensus exist. Even if we could ignore the appalling influence of extremist opinions on terrorism and wars, we would still be left with a series of problematic consequences associated with too strong beliefs. As examples, we know that some types of extreme opinions can cause serious problems in the democratic debates between the opposing political parties \cite{bafumiherron10a} or in how minorities are perceived \cite{tileaga06a}. Understanding extremism, how it starts and how it spread is obviously a very important issue.

One way that we have used to understand how extremism evolves is using the tools provided by Opinion Dynamics \cite{castellanoetal07,galametal82,galammoscovici91,sznajd00,deffuantetal00,martins08a} . Opinion Dynamics basically study how opinions spread through a society of artificial agents, whether those opinions can be classified as extreme ones or not. As such, it can help us understand the factors that contribute to a person decision to hold strong beliefs. In issues where there is actually no clear known answer, understanding the dynamics of how ideas spread and become stronger might, in principle, help us prevent problems and mayve even avoid the loss of human lives. 

Previous works in the area have studied different aspects of the question, always based on one specific ad-hoc model chosen by the authors of each paper. In purely discrete models, although there is no strength of opinion, it is still possible to introduce notions as inflexibles \cite{galam05b}. It is also possible to include a larger than two number of possible opinions to better represent not only the choices, but how strong they are \cite{bennaimetal03,nizamanietal14a}. However, strength of opinion is either not measured in those cases, or is limited to very few possible values. It is, thus, necessary to have some continuous variable that will represent the opinion about the issue we are studying if we really want to ask the question of whether one specific agent could be considered an extremist. For that reason, while discrete, Ising-like models\cite{galamjacobs07} allow us to try to understand how inflexibles agents can influence the opinion of the rest of the population, it still leaves questions answered. Therefore, there is a larger number of studies in the area that use as basis the Bounded Confidence (BC) models and their variations~\cite{deffuantetal00,hegselmannkrause02}.
One of the questions that have been investigated with the use of BC models is how agents with extreme opinions, represented by values that are close to the limits of the possible opinion values, can influence the rest of the society \cite{deffuantetal02a,weisbuchetal05}. These studies included studying the effects on the spread of extremism from factors such as the type of the network \cite{amblarddeffuant04,franksetal08a}, the uncertainty of each agent \cite{deffuant06}, the influence of mass media \cite{mckeownsheehy06a}, and the number of contacts between individuals \cite{boccara10a}. Consequences of extremism have also been studied in this context, both negative, such as escalation in intergroup conflict \cite{alizadeh14a}  as well as possible positive effects such as the possibility that extremism might help maintain pluralism \cite{gargiulomazzoni08a}.

One limitation of the BC models for studying extremism is that, in most implementations, an interaction either brings the agents to an intermediary opinion or their opinions do not change at all. In other words, unless a mechanism for the strengthening of opinions is added by hand, opinions just don't get stronger than the initially existing ones in these models. Agents with already extreme opinions have to be included in the initial conditions. On the other hand, in the Continuous Opinions and Discrete Action (CODA) model \cite{martins08a,martins12b}, it is perfectly possible to start only with agents who have moderate initial opinions and, still, due to their interactions and local reinforcement, those agents can and usually do end up with very strong views. This was accomplished by combining the notion of choice from discrete models with a continuous subjective probability each agent holds about the possibility that each given choice might be the be best one. By assigning a fixed probability that each neighbor might have chosen the best possible choice, a trivial application of the Bayes Theorem yields a simple additive model where extremism arises naturally \cite{martins08b}. 
The general idea of using Bayesian rules as basis for Opinion Dynamics has been investigated and it was observed that the generated models are actually a good description of the real observed behavior \cite{eguiluzetal15a}. And, while there is indeed strong evidence that people do not update their opinions as strongly as they should, a bias known as conservatism bias \cite{edwards68}, several of our biases can actually be explained by more detailed Bayesian reasoning, if the possibility of error in the information source is included \cite{martins06}.

By separating the internal opinion from what other neighbors observe, CODA model allows a better description of the inference process that underlies the formation of opinions. Indeed, the way CODA was built can be used as a general framework to produce different models \cite{martins12b}. These new models can be obtained simply by changing the assumptions on how agents communicate, think or use the information available to them. Indeed, traditional discrete models can be obtained as a limit case of the situation where an agent considers its own influence on the opinion of its neighbors \cite{martins13c}. It is interting to notice that the separation between internal opinion and communication is an important feature that can be included in many different models \cite{fanpedrycz15a}.  And, if the communication happens not in terms of choices but in terms of a continuous average estimate of a parameter, the BC model can also be obtained as an approximation to a Bayesian rule model \cite{martins08c}. Extensions of the CODA model were proposed to study the emergence of inflexibles \cite{martinsgalam13a}, the effects of several agents debating in groups \cite{diaoetal14a}, as well as the effect that a lack of trust between the agents can have on social agreement \cite{martins13b} and the effects of the motivation of the agents \cite{sietal10a}. Interestingly, for a completely different application, unrelated to social problems, the same ideas can even be used to obtain simple behaviors of purely physical systems, with properties such as inertia and the harmonic oscillator behavior arising from continuous time extension of Bayesian opinion models\cite{martins14a}. 
To answer the question of what factors might be more important in causing extremism, I will use the fact that the Bayesian framework behind CODA can be trivially applied to describe differences in the way the agents think and behave. That is, the framework allows us to describe both the communication process as well as the assumptions of the agents. This will make it possible to explore the circumstances where extremism is more likely to appear and make pinions become too strong. 

One first observation that can already be obtained from the literature comes from the fact that extremism usually does not develop from the interactions in BC models, while it is prevalent in the CODA model. As in BC models the agents observe a continuous value as the opinion of the agent they are interacting with, a possible explanation for the lack of increase in extremism could be associated with this different form of the communication. That is, it makes sense to ask if the difference is in the communication process.  In one case, we only communicate continuous values, signaling  doubt about the discrete choices where only the supported option is shown. And perhaps the difference in how extremism gets much stronger in CODA might be because these expressed doubts could, in principle, help prevent extremism. In order to check this possibility, a version of the CODA model where the actual probability assigned to the two options is observed instead of the preferred choice will be introduced. This will show very clearly that the solution for situations where we want to avoid extremism is not in the communication process, as the new model will lead to an actual strengthening of extremism. In this case, extremism will actually even increase faster than the original CODA model.

On the other hand, a much more subtle but important difference between CODA and BC models exist in the assumptions agents make about the problem. In BC models, agents aim to find a continuous value. It is reasonable to think they would be satisfied with an intermediary value and most of them quite often do. In CODA, the question is changed to which of two options is the best one. That means that there is a strong underlying assumption that a best answer exists. However, the ideal quest for finding the best answer might not be well described this way. It is perfectly reasonable to assume, in many cases, maybe in most cases, that the optimal choice is not one or the other, but a mixture of both, with proper weights. Examples are many. When choosing the best food among two options, the optimal solution might be take 75\% of the first one and 25\% of the second. In Politics, the optimal policy, when there are two competing ideologies, could also be to use 50\% of the ideas of each one of them. And so on.

Such compromise is not included in the CODA model nor in the variation I just mentioned. But it is central in BC models. In order to investigate this effect, a second variation of CODA is proposed here, where the agents indeed look for the correct proportion, instead of the correct option. We will see that, in this case, while strong opinions still emerge, they are nowhere near the strength of the two other versions of CODA and they might even be considered not really extreme. This result shows very clearly how damaging it can be when we simply want to believe in one idea or another, even before deciding which one, instead of looking for the best compromise. In order to compare the effects of different communicating styles with how aiming for certainty or accepting uncertainty can influence extremism, all four cases will be discussed in this paper. Two of the cases, the one with continuous communication with a decision between two certainties, and the one with discrete observation where the agents accept the whole continuous of uncertainty, are new and are actually introduced for the first time here.

\subsection{The Models}

While the first obvious difference between the BC and CODA models is what agents observe in other agents, the assumptions about the mental model agents have of the world are also subtly different. BC models assume the agents are looking for a correct value inside a range, typically between 0 and 1. In CODA, while the agents have a probabilistic estimate of the correctness of a choice (therefore, also a number between 0 and 1), they hold the belief that only one of the options can be true. In particular, by assigning a probability $p$ to $A$ being the best choice and $1-p$ to the alternative choice $B$, one is ignoring the possibility that the ideal answer could be an ideal proportion $f$ around 0.5. Instead, $p$ only represents the chance that $A$ is optimal. It makes sense to investigate what happens if agents consider the optimal would be obtained as a proportion $f$ of $A$ and a  proportion $1-f$ of $B$. While this might, at first, sound very similar to the way CODA works, we get a subtle but very important distinction. From a mathematical modeling point of view, the difference could not be clearer. Instead of a probability $p$ assigned to one specific result, the case where all mixtures are acceptable requires a continuous distribution $g(f)$ over the space of proportions $f$, where $0\leq f \leq 1$. To help the reader to know when I am referring to each case, some terminology is needed. Therefore, from now on  I will refer to agents who look for a single isolated answer as certainty wishers or, more simply, wishers. Those who accept that the best current answer can be a mixture of both choices with a frequency to be determined will be called mixers.

Of course, not only the mental model is important. If communication is continuous (and honest, as we will assume all the time here), wishers will provide the probability that $A$ is the best choice, while mixers would give us their estimate $f$ of how much of $A$ should be included in the optimal answer. But communication can also happens in a discrete way, and that can be as simple as picking a most likely choice for both agents. In this case, while using the same rule (if $p$ or $f$ are larger or smaller than 0.5), wishers are picking the option they believe more likely to be the best one, while mixers would be simply indicating which option they expect would be present more strongly in the optimal answer.
To explore the difference in communication and mental expectations, we must therefore investigate how extremism emerges in all combinations, represented by the four cases bellow:
\vspace{0.2cm}
\begin{center}
	\begin{tabular}{|p{2.5cm}|p{2.5cm}|p{3cm}|}
		\hline
		& {\bf Certainty wishers} & {\bf Mixers}  \\ \hline
		{\bf Discrete observation}	 &  CODA &  New model 2 \\ \hline
		{\bf Continuous observation}	&  New model 1 & Bounded Confidence (BC)\\
		\hline
	\end{tabular}
\end{center}
\vspace{0.2cm}
Notice that every model in this table will involve continuous variables. It is what is observed (observation) and what is expected of them (choice) that can be either discrete or continuous. The literature already shows very clearly the effects of CODA and BC models on extremism. In order to explore all these cases, we must, therefore learn how extremism evolves in the two new cases.

To make the comparison more natural, a generative framework for different models is necessary. As we have seen, extensions of the CODA model did generate such a framework\cite{martins12b} and the ideas in the framework have been used to obtain a version of the BC models that agree qualitatively with its results\cite{martins08c}. Therefore, we can simply refer to the known BC results from the literature as if they had been generated by the same framework and investigate here the consequences of the two new variations.

\subsection{Wishers with Continuous observations (New model 1)) }~\label{sec:concoda}

When we have wishers who make statements in the form of their continuous choice,  we have the case where each agent $i$ believes there is one true optimal choice, $A$ or $B$. And, instead of observing only the favorite choice of agent $j$, agent $j$ provides as information its own probability $p_j$ that $A$ is the best choice. Communication, therefore is a continuous value, just as in the BC models. The difference is that, while any probability value can be a representation of the agent point of view and communicated, there is the underlying assumption that only $p=0$ or $p=1$ (and no other value) can be the correct answers. The simplest model for this case is to assume a simple likelihood for the value $p_j$ agent $j$ communicates, if $A$ is the best choice (as there are only two choices, we only need to worry about one). For the bounded continuous variable, a natural choice is to use a Beta distribution  $Be(p_j|\alpha,\beta)$ as likelihood, where
\[
Be(p_j|\alpha,\beta)=\frac{1}{B(\alpha,\beta)}p_{j}^{\alpha-1}(1-p_{j})^{\beta-1}
\]
where $B(\alpha,\beta)$ is obtained from Gamma functions by
\[
B(\alpha,\beta)=\frac{\Gamma(\alpha)\Gamma(\beta)}{\Gamma(\alpha+\beta)}.
\] 
This choice makes the application of Bayes Theorem quite simple. Assuming agent $j$ information has some value, it is to be expected that $\alpha$ and $\beta$ are such that $E[Be(p_j|\alpha,\beta)]>0.5$.  This translates, in the likelihood to the condition that $\alpha>\beta$. Applying Bayes Theorem and using the same transformation to logodds $\nu_i$ as in the CODA model
\[
\nu_i=\ln \left( \frac{p_i}{1-p_i}  \right),
\]
we obtain the very simple dynamics
\begin{equation}\label{eq:logoddcontinuouscodanu}
\nu_i(t+1)=\nu_i(t)+(\alpha-\beta)\nu_j .
\end{equation}

\begin{figure}[ht]
	\includegraphics[width=0.95\textwidth]{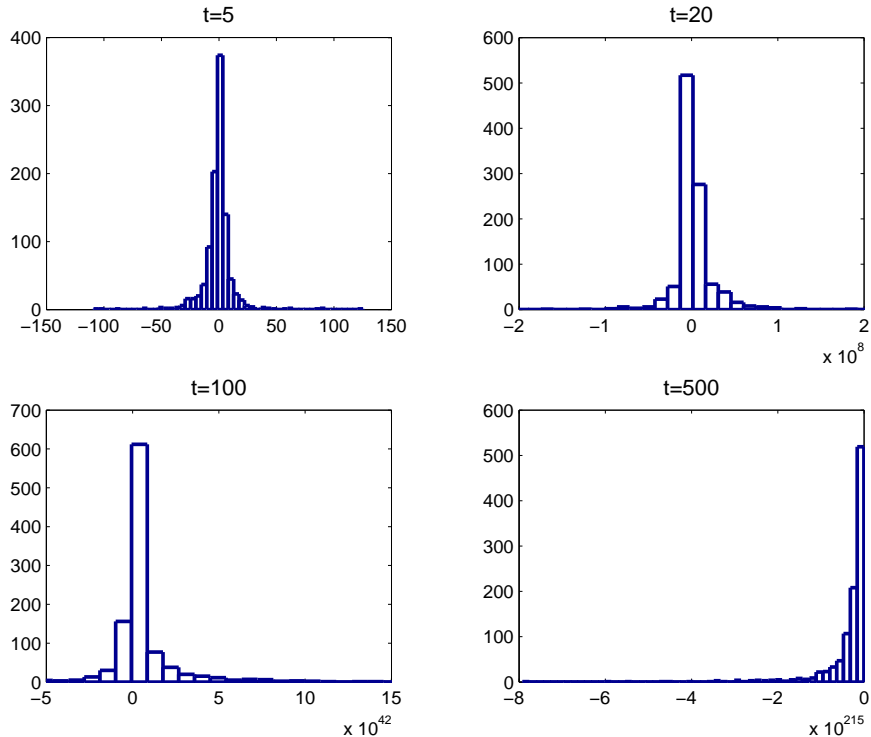}
	\caption{Distribution of the logodd opinions $\nu_i$ after $t$ interactions per agent.}\label{fig:contcodanlat32distr}
\end{figure}

This is similar to the CODA dynamics (where we would have $\nu_i(t+1)=\nu_i(t)+\delta$, with a fixed $\delta$), except that now, at each step, instead of adding a term that is constant in size, we add a term that is proportional to the log-oods $\nu_j$ of the opinion of the neighbor $j$. This means that the system should also tend to the extreme values, in opposition to the BC models where verbalization was also continuous but the agents were mixers. Here, as $\nu_j$ increases, so will the change in each step, making the long term increase in $\nu_i$ no longer linear.  Simulations were prepared to confirm the full effect of this change when each agent only interacts with  a fixed neighborhood. Square lattices with periodic boundary conditions and von Neumann neighborhood were used and the state of the system observed after different average number of interactions $t$ per agent. 
During the initial stages, we observe a behavior very similar to that of the CODA model, with a clear appearance of domains for both possible choices. However, instead of freezing, those domains keep changing and expanding and, eventually, one of the options emerge as victorious and the system arrives at a consensus.

Figure \ref{fig:contcodanlat32distr} shows histograms with the observed distributions for  $\nu_i$. The important thing to notice there is how the scale of typical values for $\nu_i$ changes with time. In the original CODA model, the extremist peaks, after the same number of average interactions per agent, corresponded to  $\nu_i$  around $430$ (that is around 500 steps away from changing opinions, with the size of the step determined by $a=70\%$). That value was already quite extreme since it corresponded to probabilities of around $10^-{300}$. Now, however, instead of peaks of $\nu_i$  around $-430$, we can see at \ref{fig:contcodanlat32distr} we reach $\nu_i$ values in the order of  $10^{215}$, corresponding to the exponential increase in $\nu_i$ predicted by  Equation \ref{eq:logoddcontinuouscodanu}.

Indeed, the system as a whole observes a surprisingly fast appearance of extremists. After as little as 5 average interactions per agents, we have opinions so extreme that it would have taken 100 CODA model interactions to obtain. The lack of freezing in the domains is actually due to these ever increasing opinions. As one side domain randomly gets a much larger value of $\nu_i$, this strength is enough to move the border freely and eventually, one of the options wins.

\subsection{Mixers with Discrete Observations (New Model 2)}~\label{sec:conchoicecoda}

The last remaining model to study is the opposite of the previous one. Now, each agent only states its best bet on which of $A$ or $B$ should make a larger part of the optimal answer. Unlike CODA or the previous model, instead of trying to determine which option is the best, the agent is trying to find the correct mixture of both choices. That is, we can no longer talk about a probability $p_i$ agent $i$ assigns to $A$ being the optimal choice (and $1-p_i$ for $B$). Instead, we need to have a continuous probability distribution $g_i(f)$ over the values $0\leq f \leq 1$ for each agent $i$ and a rule for the agent to make its choice based on this distribution.  The simplest rule is that the agent communicates its best guess, based on whether the expected value of $f$ is larger or smaller than $0.5$.

As the agent tries to determine the optimal frequency $f_i$, this problem can be seen as a very simple problem of inference on the probability of a Binomial random draw. Once again, the $Be(f_j|\alpha,\beta)$ distribution is useful, except this time, it is the natural choice for the prior opinion $g_i(f)$, since it is the conjugate distribution to the Binomial likelihood. The values of $\alpha_i$ and $\beta_i$ now represent the belief of the agent $i$ and the average estimate is very easily obtained from $\frac{\alpha_i}{\alpha_i+\beta_i}$. That basically means the agent $j$ states it thinks a larger proportion of $A$ is preferable if $\alpha_j>\beta_j$, and a larger proportion of $B$ corresponds to  $\alpha_j<\beta_j$.

The update rule one obtains is also trivially simple. If $i$ observes $j$ has a preference for $A$, it alters its distribution by making
\[
\alpha_i(t+1) = \alpha_i(t)+1
\]
and it keeps $\beta_i$ constant. If it observes $j$ chooses $B$, then $\alpha_i$ is unchanged and we have
\[
\beta_i(t+1) = \beta_i(t)+1
\]

What is interesting here is that, while the values of the internal probability can be quite different, the qualitative rule for when the opinion of one agent changes from $A$ to $B$ is exactly the same as in the CODA model. Indeed, the above rule is equivalent to adding or subtracting one to $\alpha-\beta$, depending on whether $A$ or $B$ is observed. And the opinion changes when  $\alpha-\beta$ changes sign. That is the exact rule of the CODA model, except its dynamics was based on the logodds $\nu$. As the actual dynamics of $\nu$ could be renormalized to increments of size 1, both models will yield, given the same initial conditions and same random number generator, the exact same path in terms of configuration of choices on the lattice. On the other hand, in the original CODA, probabilities change much faster. For example, if after 500 interactions, one specific agent had  made 465 observations favoring $A$ and 35 favoring $B$, we have seen that the the effect of these 430 total steps for $A$ meant the probability $1-p$ of $B$ being the best option  goes all the way down to $10^-{300}$. Here, if we start with $\alpha \approx  \beta \approx  1$ (a choice close to uncertainty), we will have an average proportion for $A$ of $466/502$, something close to $93\%$. From this point, it would take the same number of influences in favor of $B$ to change agent $i$ decision towards $B$ as it would have taken in the original CODA, that is 430 interactions favoring $B$. The opinion on which side should be more important is equally hard to change. But the very strong extremism we had observed there is gone, replaced by a reasonable estimate that the optimal proportion should still have a little of $B$ ($7\%$).

\subsection{Communication or Assumptions: Comparing the Four Models}

\begin{figure}
	\centering
	\includegraphics[width=0.95\textwidth]{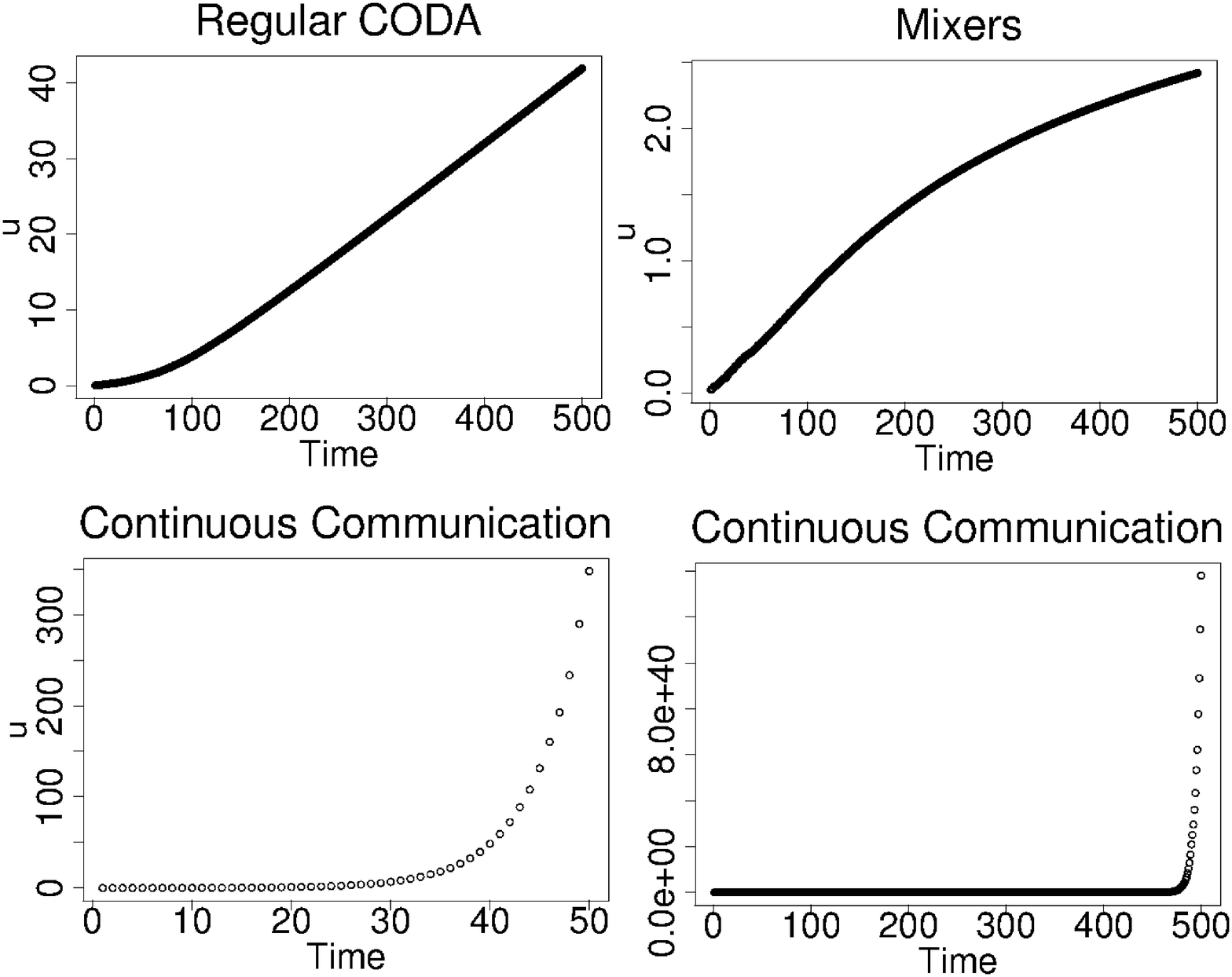} 
	\caption{Evolution of the average value of strength of the opinion, measured by $\nu=\frac{p}{1-p}$ as a function of time. Time is measured in average interactions per agent and each point corresponds to the average value of 20 realizations of the problem. Due to the extreme difference in values, the evolution could not be visually shown in the same graphic. The upper left graphic corresponds to temporal evolution of a population of wishers with discrete communication, that is, the regular CODA model; the upper right, we have mixers with with discrete communication (new model 2); while both graphics in the lower line show the case of wishers with continuous communication (new model 1), at left, just its evolution to 1/10 of the time the previous cases evolved, at right, its evolution for the whole range of times.}\label{fig:comparison}
\end{figure}

To make comparing the four cases easier, I have run a series of simulations where every agent can interact with all the others, randomly drawing an agent $j$ to observe at each interaction. That means all the systems will eventually reach consensus. The idea is to avoid any consequences of locality and focus only on how the strength of the opinions evolve. In all cases, the initial conditions included half of the agents showing some form of preference for each case. The magnitude of the difference in the evolution of the opinions can be better grasped in the graphics shown in Figure \ref{fig:comparison}. For the case where we have wishers and discrete communication, the strength of opinion increases so fast that it is simply impossible to show its evolution in the same graphic as the other two cases. If we get the other two cases in the same graphic in a way that their evolution can be observed, the wishers with discrete communication soon move out of range. Therefore, each case is shown in its own graphic. In order to understand the graphic better, note that odds of 1 to 10 translate to values of $\nu=2.3$; 1 to 100 corresponds to $\nu=4.6$; 1 to 1,000 is $\nu=6.9$, and so on.

What we see is that the difference between wishers and mixers is probably a crucial key to prevent extremism. Interestingly, for mixers, when they observe choices and not the internal probability, that still allows them to get opinions that do become stronger with time. When they do observe the continuous evaluations, we have the results of the Bounded Confidence models, even if we do a Bayesian version of the problem \cite{martins08c}. And, in this case, if no trust considerations are introduced, as it is the case of the other results here, the average opinion $\nu$ will always stay around zero ($p=0.5$) for an initially random population, the exact opposite of extremism. That is, observing the amount of existing uncertainty make mixers estimate the actual proportion of support for each idea in the population. When they only state their preferences, random effects, even if small, do make one of those preferences obtain a little more support and, as a consequence, the perceived optimal mixture shifts to favoring that one random choice. But we still never get to real extreme values. 

The exact opposite effect is observed for wishers. In this case, just observing the choice of their neighbors make their opinions already strong enough to be described as very extreme. While we might have expected that observing the uncertainty or, in other terms, doubt on the part of the other agents would lead them to some amount of doubt and, therefore, less extremism, this was not the case at all. The effect we observed for wishers can be described better in two steps. Observing doubt just once could mean a smaller change in the probability the agent assigns to the best option (this can be easily adjusted by the model parameters). But in the long run, the opinions get stronger as constant reinforcement happens. Even if we start the system with smaller increases, as the agents observe their peers becoming more confident, their opinions start changing faster. It makes sense, as people with stronger opinions are expected to have a stronger influence. And that is where the problem starts because there is no limit on how strong those influences can become. At some point, the change in opinions will become as strong as it would have been if only the choice had been observed. From there, opinions will still keep increasing, making those changes larger and larger with each interaction. This influence actually grows exponentially and, given time, we can observe the absurd increase in certainty we see in Figure \ref{fig:comparison}.

The message from these models is clear. Mixers opinions seem to never grow as strongly as the those of wishers. Indeed, either they will not grow at all, as in Bounded Confidence models, or they will approach certainty (a mixture with frequencies of $f=1$ or $0$) due to random effects causing one option to receive a greater amount of support. But even then, we only observe $\nu$ growing logarithmically. On the other hand, wishers always observe a strong growth of the probability as shown by the linear or exponential growth of $\nu$, depending on how communication happens. Surprisingly, uncertainty in communication can even make their opinions much stronger in the longer run, even if it might have had some short term benefits. And all of these results were obtained without even considering the very real problem that, as a person opinion becomes stronger, she will probably start disregarding opinions that disagree with her own. 

Those results bring some remarkable evidence on how to deal with extreme opinions. Changing the way we communicate might not prevent extreme views at all, and if we do not consider long term effects, it might even make things worse in the long run. We should notice that, in the cases studied here, the strengthening of the opinions happened only by social influence, no observations of the world were ever made. That means that there was no real reason for the system as a whole to go to such extremes as there was no real data confirming either possibility. The problem with unwarranted extreme opinions seems to be deeply connected with the way we think. It is a consequence of our wish to find one truth, of our desire to believe and take the side of one option, even in cases where, at the beginning, every agent had mild and easily changeable positions.

The message this exercise in Opinion Dynamics adds to this paper conclusions is clear. Not only our self-defining beliefs make us unable to think correctly and consider all evidence as well as not being supported by any kind of rational thought. This tendency to pick sides also make our opinions absurdly stronger than they should be. This tendency seems to be at the very heart of why extremism arises. The agents in these simulations were willing to change their opinions given enough influence to show them wrong. Even under these idealized conditions, the simple desire to have one option as the best answer already caused severe problems, making opinions absurdly and unwarrantedly strong. This only reinforces the evidence we had, from Cognitive Psychology as well as both Deductive and Inductive Logic, that we should not take sides. Indeed, that conclusion just became stronger: we should not even want to take sides. In the next Section, we leave the study of extremism behind and I will inspect the consequences of our wrong tendency to belief by discussing the effects this bias causes on the ways we make Science.

\section{Scientific choices, beliefs, and problems}\label{sec:science}

	\subsection{Creating and Evaluating Theories}
	
	The na\"ive description of the scientific enterprise, the one to which most of the public and most of the scientists probably subscribe to is that it is the job of Science to find the one truth about the world. Often, scientists actually believe they have done just that. And we have learned earlier here that this will make them unreliable sources about the ideas they defend. But even when they recognize that we still are far from knowing it all, that our current ideas might be the best ones we have but are only tentative descriptions, they still often expect that one day we will create that one winning theory, the Grand Unification that will explain everything. That might come to pass, of course. But we are nowhere near it and, given all we have seen so far, I am forced to question how much of our human failures might be influencing the reliability of the Science we do today. And, if there are problems, what can we do to mitigate them.

	The desire to have one winning theory, the one idea that will rule them all, the assumption that it is possible to find that idea, as we have seen, ought to have consequences. When using inductive methods, it might actually be the case that a mixture of the available theories does a much better job than each one individual idea can do by itself. And, in any case, ideas are to be ranked by how probable they are, not discarded nor can we prove them right. They do not win or lose in absolute senses, even if they can still get arbitrarily close to that. We have just seen in Section \ref{sec:extreme} that when we look for the one correct idea, our estimates can become too strong, stronger than they should be. And that considering mixtures can actually lead to much weaker, less extreme opinions, even when all agents end up agreeing with each other. Our opinions should  probably be quite less strong than they are, especially in the cases when scientists engage in hot debates about who is right, assuming that the truth must be with just one group of them. In some rare cases, when there is strong evidence that neither idea does a perfect work, we actually have seen the survival and acceptance of a mixture of even incompatible theories. Such is the case of Quantum Mechanics and General Relativity, as each of them describes a different set of observations with what is considered remarkable accuracy and fail at explaining the phenomena the other theory describes. But this reasonably peaceful cohabitation of the minds of the researchers seems to be a rare case, forced by the utter failure of each theory to provide decent predictions for the set of experiments where the other one meets its greatest successes. And also for the amazing successes in predictions that both theories have met in their own field of applications. It would be good to look at this example and think that the scientists are working as they should, but they actually never had a choice under this combination of circumstances. And physicists still look for the one theory that will unify them both, as they should do. The amazing success of physical theories is both a very luck case that allows physicists to ignore most considerations about mixture of ideas and probabilistic induction (we will see why bellow) and also a curse that has prevented them from learning those exact same lessons. And that means they still believe in the eventual blind acceptance of the one theory that might never come. And, even if it does, our current logical apparatus does not allow us to be certain, even if we actually had the exact correct description of the Universe. That is, as far as we know, an unavoidable human limitation. 
	
	When we think about how scientists should behave, we realize we have always had idealized prescriptions about how the nature and demands of the scientific work. Standards are extremely important, of course. And, given the enormous success of the scientific enterprise, it used to make sense to assume that individual scientists did not deviate much from the norms of the area. Some of them certainly did, but errors and deceptions ought to be the exception, right?  However, since the first observations of how scientists actually work with their ideas, it has become very clear that we do not follow the norms people believed we did. And that happened even among people working in some of our greatest technical feats, such as the Apollo missions \cite{mitroff74a}. The departure of the idealized norm seemed to be so prevalent that the personnel in those missions considered normal that people they kept respecting intellectually could hold dogmatic positions and not be convinced of their erroneous views even by data. Scientists and high-level technicians are not different from the rest of mankind and they also defend their ideas. And that should actually  be no surprise to anyone working in a scientific area.
	
	Feyerabend\cite{feyerabend94a} observed, from historical cases as well as his own interactions with working scientists, that the typical behavior did not match the ideas he had been told were the right way to do Science. His observations were quite influential and have even led people to think the whole reliability of the scientific enterprise could be seriously compromised. A shortened version of the rules he investigated if we actually obeyed are the tentative norms that scientists should not use ad hoc hypotheses, nor try ideas that contradict reliable experimental results, nor use hypotheses that diminished the content of other empirically adequate hypotheses, neither propose inconsistent hypothesis. And yet, he found evidence that all these norms that, at first sight did sound reasonable, were broken on a daily basis. More than that, important advances had been achieved exactly because those rules had been broken. If those rules were indeed the source of Science credibility, it would have looked as if there was no longer any reason at all to place more trust in scientific knowledge than in any other form of knowledge.

	But that is not really the case. One of his main conclusions, that all ideas are worth pursuing, is, quite contrary to what he believe, actually a prescription of a number of inductive methods. Instead of limiting the ideas that should be tested, inductive methods actually support the exact concept that every idea must be kept, tested as a possibility, and ranked. And, once again the fact that we are limited plays an important role. We can not know every possible idea and demonstration in advance, or ever. For us, limited humans, the need to test all ideas actually means that we must work to get more and more ideas created. Only that way we can add them to our space of possible descriptions. Most of them will be utter failures, with probabilities so close to zero that they will be indeed useless. But we only know that after we come up with them and verify if they match the world we live in. 
	
	If we want real probabilities and not only odd ratios between two ideas, Bayseian methods actually require, for the ideal situation, that we should consider all possible explanations \cite{fitelsonthomason08}. Similarly, Solomnoff method for induction \cite{solomonoff64a} requires a theoretical computer that would be able to generate every possible program, up to infinite length, and average over all of programs that produce as output the data we already know about a problem. While both approaches are obviously impossible, they do provide the ideal situation. As we work to get closer to the ideal, even if we will never reach, we have to realize that we must indeed catalog all possible ideas, including those very same hypothesis Feyerabend had been taught to be forbidden. Obviously, hypotheses that contain logical inconsistencies are not acceptable per se. But even those can be useful as a middle step, if they inspire us to later obtain a consistent set of hypotheses, by correcting initial  problems and inconsistencies.
	
	However, while generating the largest number of hypothesis is something we should strive for, this prescription does not lead to any kind of relativism at all, contrary to what Feyerabend's followers would like to believe. We just need them all to verify, as well as possible, which ideas actually match our observations better. Not defending an idea means not disregarding any possible theories as a possibility. But it is unavoidable that some of those theories will turn out to describe the data so badly that, while their probabilities will not be technically zero, they will so close to zero that we can safely disregard them when making any statements. That is different from proving an idea wrong, from an epistemic point of view, but the practical consequences can be exactly the same. In Science, the difference between proving an idea wrong and estimating its very low probability is often just a mathematical distinction, with no real consequences. As we have limited time and resources, too unlikely hypotheses really ought to be disregarded as if they had been proven wrong.
	
	This poses the question of how much we can get close to prove a theory ``wrong''. The answer can be illustrated by the classical observation of the advance of Mercury's perihelion and why this experiment was seen as a ``proof'' that Newtonian Mechanics had to be replaced by General Relativity. One important thing to notice before we start is that we are talking about two theories that agree incredibly well with experiments, that have very close predictions in many circumstances and that, as far as we know, provide predictions, in those cases, that match experiments with amazing accuracy. However, there are indeed cases where they disagree in their predictions and, as those cases started to be tested, there was no longer any doubts in the minds of the physicists about which theory was superior. The subtle differences in where Mercury was when it got closer to the Sun and how that varied over centuries was one of the first very strong ``proofs'' that General Relativity was indeed better. So, why can physicists use the term proof in these cases, even when no real proof exists.
	
	What people knew when the two theories started competing was that when we estimate every influence of the other planets on Mercury's orbit using Newtonian Mechanics, its perihelion should move $5577.18\pm 0.85$  seconds of arc per century. However, the actual observed displacement was $5599.74\pm 0.41$ \cite{clemence47a}. On the other hand, General Relativity introduced a correction of $43.03$ to the Newtonian prediction, bringing the actual estimate of the difference between observed measurement and prediction from $42.96\pm 0.94$ to $0.07\pm 0.94$ (the errors have been estimated by me simply adding the variances in Clemence article and are, therefore, not much more than an educated guess). We have the Newtonian prediction at a distance of 45.7 times the standard deviation from the observation and the General Relativity at a distance of 0.07 standard deviations. Therefore, if we assume the observed variations have a Normal distribution, the likelihood ratio between the two theories will be
	\[
	L=\frac{e^{-\frac{-0.07^2}{2}}}{e^{-\frac{-45.7^2}{2}}}=\frac{0.9976}{3.1\times 10^{-454}}=3.2\times 10^{453}.
	\]
	In other words, if you started by believing General Relativity had just one in a billion chance to be a better description (1 in $10^9$), you would end concluding it was actually better by a chance around 1 in $10^{453-9}=10^{444}$.
	It is very important to notice that this calculation is heavily dependent on the assumption that the errors would follow a Normal distribution, which is a rather thin-tailed distribution (very distant values are extremely unlikely). If we do change the distribution of the experimental errors to a distribution with much larger tails, as for example, an arbitrary t-distribution with 10 degrees of freedom, things change dramatically. In this case, the likelihood ratio would be now
	\[
	L= \frac{\left( 1+\frac{0.07^2}{10} \right)^{-11/2}}{\left( 1+\frac{45.7^2}{10} \right)^{-11/2}}=\frac{0.593}{1.7\times 10^{-13}}=3.5\times 10^{12}.
	\]
	Now, an initial opinion of one in a billion against General Relativity would be transformed in just 1 in 1,000 in favor of it. However, the change in opinion is still quite impressive. And we must remember this was just one single experiment. Since then, many other experiments have confirmed the predictions of General Relativity. And we must now include the results of all these experiments to obtain a final estimate. It is not difficult to see that the probability of Newtonian Mechanics being correct is basically zero for all practical purposes. Results in Physics have been so discriminating between competing theories in most cases that physicists can work perfectly well if they assume they are proofs, even though they are not real logical or mathematical proofs.

	One thing to learn from the difference in the behavior of the likelihood ratios is that the results can indeed depend heavily on the assumptions we make about the nature of the experimental errors. Including both possible hypothesis on the errors, along with many others, as we should, can complicate these raw estimates quite severely and that is beyond our objectives in this article. Indeed, some hypotheses about the errors will be more probable and that has to be taken into account and we are soon in a deep sea of technical statistical problems. For out purposes here, we just need to observe that a theory can really become extremely unlikely to be true given data. For both calculations, the change in probability was actually quite strong. Technically, that did not mean Newtonian Mechanics probability was a mathematical zero. But we can say that it is so ridiculously close to zero that saying it in any other way is really a waste of effort, unless you are indeed a mathematician.
	
	Notice also that this exercise was a comparison between only two theories. Ideally, we should keep all possible alternatives at hand. But the vast majority of other alternatives describe the data so much worse than the ``wrong'' Newtonian Mechanics that we do not even bother to enumerate them. As I have mentioned above, physicists have been quite lucky in the way data can discriminate between competing ideas. When that happens,  concerns about our limited reasoning power are far more relevant than the prescription that all ideas should be kept. Actual calculations of most possibilities can easily become too complicated to be carried out if more and more hypothesis about the experiments are introduced as well as more theories about the Mercury's orbit and there is nothing to gain from those. It does make perfect practical sense to just keep those theories that seem reasonable enough at the moment.

	On the other hand, Feyerabend conclusion that we must keep generating new ideas, no matter how well they initially match what we know right now, was actually normatively correct. This has some interesting consequences that people seldom consider. One of them is the actual strong usefulness of the existence of purely theoretical scientists. They are already a major part of the culture in the physical sciences and they also exist in some areas of Biology, such as Evolution. Having people dedicated solely to coming up with new better ideas means we can generate those needed ideas faster. Advances can appear faster and the areas where they exist are bound to benefit from their specialized knoweldge.

	There is also incidental and very strong benefits from having a large number of people who do not make experiments or generate any data. And that is the simple fact that the people who will check their ideas are not the same as the proponents of the theories, allowing for a much less personally-biased analysis. The current tendency to demand that people produce the ideas and the data that supports those ideas can result into biased choices of sets of data. Data that comes together with a new idea is seldom reliable at all. There are reasons for that to happen related to how some areas do bad statistical analysis and bad practices for accepting papers, but we should also not ignore the fact that human reasoning is deeply flawed. Even while trying to be honest, scientists will suffer from confirmation biases and their reasoning will probably try to deceive them into creating arguments that favor the ideas they support. Someone proposing the idea and providing the tests for it is actually a very strong moral hazzard that we should try to diminish. 	
	While scientists might not be used to the idea of moral hazzards, we are not immune to them. We really should build our areas in a way that makes it less easy to happen. Having pure theorists is not only a good idea about dividing the work and generating needed new ideas faster. It is also a good idea because it avoids ethical problems in data analysis, as well as human biases and our tendency to defend our own ideas. Of course,  the existence of theorists require that we know enough about a specific problem that we can start generating several theories and this might not be the case yet for some areas of knowledge. Still, having a number of individuals solely dedicated to pure theoretical work and some exclusively dedicated to experimental issues is quite likely something each area should strive for, even if that might not be possible today in every field of knowledge.

	Another incidental and very important benefit from having theorists comes exactly from all our reasoning failures. Expecting our brains and our arguments to help us arrive at conclusions we can trust is, at the very best, a foolhardy proposition, as it should be clear now.  On the other hand, a recent study on judicial rulings has shown that it seems that legal training and experience can provide some resistance to identity-protective cognition. Unfortunately,that resistance only applied to the situations about legal reasoning for which the judges had been trained\cite{kahanetall15a}. This suggests that it is possible to train people to make decisions that are close to neutrality, at least on matters that are clear, logical applications of already known ideas. It seems to me that a central aspect of this training effect might come as a consequence that legal training is indeed based on a series of well defined rules and, therefore, its members can learn to apply those rules and avoid normal human biases. This seems no different from the fact that, while we naturally fail miserably at simple logical problems, we are very good at solving them when they happen in an everyday situation.
	
	 We should also remember that there are several experiments that show that professionals actually fail at their own tasks and at evaluating their own rate of success when they do not have access to fact-correcting strategies. This highlights the fact that simple training alone is not enough. We need training, we need to learn when we are making mistakes and, whenever possible, we need to making thinking tools to correct our human nature. Mathematics and Logic play exactly that role. And it is a role that is probably similar to the legal training that judges receive and allow them to identify which cases should be treated in which way. For scientific knowledge, Mathematics plays exactly the role of thinking in the sciences where we have been able to successfully apply it. A mathematical or logical demonstration is not subject to our biases, since anyone with the proper training will find mistakes that are not too subtle and hard and expose any errors. It is not Mathematics that is unreasonably effective \cite{wigner60a}, it is the standard for comparison, our own reasoning, that is far poorer than we would like to admit. This importance of having logically sound reasoning in every field of knowledge makes the existence of classes of scientists specialized in applying mathematical tools to the exploration of each area another very strong defense against human failings. In particular, given the complexity of the task, theorists are a necessity in this case, since it is not reasonable to expect one single individual to master the intricacies of a fully developed area and its theories as well as the many difficulties associated with data gathering and analysis.

	The need for the creation and acceptance of new ideas as part of the theoretical repertoire of an area has at least another important consequence, probably far more important than the ones I have described above. While individuals might have difficulty accepting ideas that contradict their favorite theories, we must make it sure that this difficulty must not happen in any area as a whole. Cultural norms can help offset the bad behavior of the scientists \cite{merton79a}, but we do need to verify that there are social effects in each area that make the knowledge generated by the area researchers more reliable. In every field, new ideas should be always welcome or, at the very least, not forbidden, even though most of them might prove to be very improbable. Unfortunately, the needed openness to new theories might not be a common characteristic to all areas. Heavy reliance on the use of authority figures, a need to show that someone has said something similar before, these all too common practices can lead whole areas to stop advancing. If agreement with what has already been said is expected, even if researchers were not prone to human biases, we would expect advance to come to a halt. Add the fact that we are all humans and will defend our old positions and we are starting to head to a nightmare scenario. And, while this problem seems to be more serious in the Humanities, it can actually be observed everywhere. The refuse of many physicists to accept String Theory as a theoretical construct on the basis that it does not provide new predictions is epistemologically absurd, even more when we consider that its predictions actually work, since the identical predictions that already existed work. The problem many people see with String Theory is that it just provides, for the experiments we can do, the same predictions of the older models. This goes directly against the belief physicists have that there will be only one theory. And, as humans, they feel the urge to defend that random belief. Ironically, from an inductive point of view, the truth is that having more than one theory providing the same predictions is actually epistemologically useful. This is easier to see using Solomonoff framework. There, more identical predictions generated by different program makes those exact predictions more probable \cite{solomonoff64a}! And that actually means that the predictions of those theories in Physics become a little more reliable, which should be considered a good thing. But String Theory does poses a problem to the side many physicists chose to take on philosophical and metaphysical issues.

	Finally, it is worth noticing that we already have some simulation evidence that, if social influence becomes too strong, an area might fail to replace an old theory for a better one \cite{martins10a,martins13a}. The habit of constantly referring and excessive citing of old works is, indeed, something that should be considered at least very worrisome. It basically just reinforces the already existing ideas and this tradition can pose a strong barrier to the appearance of new theories. We need to allow for those new theories to be created and developed to the point they can be compared to reality. It does not matter if their predictions are not new or if we might not like them or if they may sound too weird at start nor if they go against what is the consensus in a given field. Competence is always a need, of course, but that is the only requirement of the scientific game. A good part of what makes an area trustworthy might be just its ability to allow the appearance of the new, the creation of ever new competitors for the position of the best description we currently have.

	\subsection{Tests and the Desire to Know}

	Of course, not only openness to new ideas matter. Simple competence is of fundamental importance, and one aspect of being competent is our competence in data analysis. This is paramount to any reliability of the conclusions. But the history of the use of statistical methods in areas as Medicine and Psychology is one where competence was traded for ease of use. And, in this case, the ease of use was also deeply associated with our desire to find ideas to accept and defend. While statistical tools have been developed to points where many of them might not be described as easy, many of those tools have been created as statisticians were prey to our human tendencies. Give me an idea to defend is a recurrent theme in the development of the area. And that had to lead to a new set of problems.
	
	Actually, things have become so out of control that there is good reason to consider most results of these areas as unreliable \cite{ioannidis05a}. The causes are not only statistical methods but also the lack of deeper statistical knowledge from professionals in the areas where the problem is more severe, coupled to absolutely damaging politics from the journals about when to accept an experiment paper and when not to. Still, it is clear now that the use of methods that reject ideas is at the very heart of the observed problems. Once more, I have to repeat that we need all ideas.  And since we have serious psychological biases when we decide in favor of a specific idea, accepting and rejecting theories or hypotheses is a very bad way to do Science, even if we did not know that. Once again, it is worth remembering that, during practical applications, a choice might be needed. A physician, after diagnosing a patient, might have a number of illnesses that fit the data and different, incompatible treatments for each disease. In this case, not only the probability that each diagnosis is the right one should be considered but also the severity of the consequences of each choice. A full decision procedure is needed and that will often mean, in practical terms, to behave as if one of the possibilities were the true one. But this is true only for the decision about which medicine to prescribe. The physician must actually keep all the alternatives as possible explanations in mind and update the probabilities as the case advances and more observations become available. This can mean changing the initial decision as more is learned. Decisions can ignore possibilities, the knowledge states of our minds must not.

	That means that any data analysis that wants to avoid generating biases in the scientists and readers minds should not include statements that could sound as if we were pointing at one idea as the correct one nor at one idea as the wrong one. More specifically, we should avoid any kind of hypothesis testing, regardless of its statistical basis. Instead, ideally, we should just point to which ideas are or seem more probable or, when that is not feasible just present and provide useful summaries of the data. An article that estimates rough probabilities is not a problem, if it is clear the estimates are rough.  The same is valid for an article that only shows the characteristics of the data and how it varies. Strong statements with no very clear estimate of uncertainty are problematic. This prescription is something different researchers have been obtaining as they study the many errors associated with bad statistics and how it interacts with publishing practices. That the absurdly disastrous practice of some areas where most journals would only publish results with ``strong'' $p-$values has led to nefarious consequences due to publication bias is something we already know  \cite{ioannidis05a}. But more recently, we have observed many scientists also pointing the serious problems that arise from making tests, regardless of whether the journals would publish all experiments or just a severely biased collection of them.

	The good news is that the misuse of several statistical tools and the consequences of such misuse have finally been described as a serious problem\cite{nuzzo14a}. In Medicine, cautionary articles about this misuse exist for more than two decades \cite{altman94a}. The widespread, unthinking, and stupid practice of using $p$-values as a way to avoid thinking about the statistical problem has been correctly labeled as surrogate science \cite{gigerenzermarewski15a}. Warnings about how $p$-values can be useless have been already published \cite{cumming08a}. In order to still be able to use the existing reported $p$-values in the already existing literature in a non-testing manner, some correcting techniques have been proposed \cite{held10a}. Unfortunately, while they make sense from a statistical point of view, they can not correct for the fact that the reported values we have in print are useless, as they are also the consequence of the publication biases.
	Sadly, the decade old conclusion that most published results are false in the areas that make widespread use of these tools \cite{ioannidis05b} seems to be still a valid conclusion today.	
	The need to completely abandon null-hypothesis significance testing has been already acknowledged \cite{cumming14a} but the suggestion to simply change it for confidence (or credibility) intervals is not necessarily an improvement from the point of view of the reader of the published information who might still read it as strong support to one idea \cite{savaleidunn15a}. On the other hand, acknowledgment of the dangers of null-hypothesis significance testing have finally reached the point where the method has simply been banned from at least one journal \cite{trafimowmarks15a}, so things are improving, albeit slowly. It is clear however that the misuse of Statistics in the literature is not limited to problems with tests and $p$-values and represent a much larger problem with competence in general \cite{leekpeng15a}. But tests do sound too much as taking sides. As such, you should not allow humans to use them, under penalty of obtaining unreliable analysis. Either from the statisticians or from the readers of those analysis.
	
	For all we have learned here, I must suggest that an important part of the how we can correct our errors is to understand the evils of our desire for possibly non-existent final answers. That desire is why hypothesis tests were created in the first place, to rule out ideas, despite the ever obvious fact that this ruling out never meant that we knew which hypothesis was true. But the terminology seemed to imply we did, with rejection or non-rejection, as well as acceptance of ideas the normal way people mention those statistical results. While the only truth is that data only makes some ideas more probable and others less probable, with cases where we can indeed stop worrying, when probabilities actually go down to basically zero. Estimating the actual probabilities might be too difficult for many problems. How to do it is often an open problem for researchers in Statistics, but the fact remains that choosing sides about hypothesis is most of the time completely unwarranted. Researchers must understand as soon as possible that statistical techniques do not provide magical answers; they will just point at more probable answers, not at the true ones. Epistemologically, no idea should ever be rejected.

\section{Discussion}

Changing from normal, committed beliefs to probabilistic, uncertain beliefs might not be easily done. Our societies and our institutions are currently built around the concept that ideas should be defended. We teach our kids to stand for what they believe, we teach them that debating means defending instead of analyzing ideas.  But it should be clear now that the fact that we have always done that does not mean this is the best way to do things. What we have seen here strongly supports the idea that we should start learning as soon as possible that ideas are to be challenged, never supported. They can be often safely discarded, as probabilities fly to zero. But openness to change is paramount. This is a concept that scientists have always claimed to follow, even though we do not do that with our own theories. It might be useful to distinguish in our language the two types of beliefs, supporting unwarranted beliefs from probabilistic ones, to make it clear that they are not the same. They do not have the same effects, they are far distant in their consequence. Probabilistic beliefs are correct while committed beliefs make us stupid.

Probabilistic beliefs (peliefs?) are normatively acceptable. By claiming a probability we are not committing ourselves to an idea, at least not so strongly. It becomes much easier later to accept those ideas were wrong. The consequence of this simple fact is almost obvious and yet, almost unsaid. Socrates sounds like he had grasped it, when he used to claim he had no knowledge about the world, but we have kept making this same mistake throughout the millennia since then. We should not teach anyone to debate and support an idea, we should teach people to look for mistakes in all the sides of the discussion. We should not think that one political side will have the right ideas and should be supported against the wrong side. It is perfectly possible and sometimes very likely that one of the sides will be correct more often than the other side. But chances are that good and bad ideas will come from both sides. 

We defend points of view in our work, regardless of the field of activity, when we have no valid reason to expect that we would be always right. Accepting that the ideas we consider probable do not define who we are might be an important step. The things I believe about how the world is are just my best guess. Sometimes a very well educated guess, sometimes a wild guess, but they are always just a probabilistic guess. It is fundamental to acknowledge that they might just be wrong. At best, it is my fallible opinion that they are likely to be the current best option. And we all must admit that we will often be very wrong. Our human nature will make that a certainty.

There will be cases when taking sides will be unavoidable. Moral issues might be an important exception to the rule of never taking sides. We do not know there is a right moral answer and, while the vast majority of us will agree on many basic ideas, it is not clear that there are right answers. Much less so for those ideas where we disagree. Defending your own moral values might not be a problem, assuming you do know them perfectly well and that they do not suffer from logical inconsistencies. However, a significant part of the political and economical discussion is not about morals, it is about which programs will work better. When that happens, taking sides will only prevent us from learning. It is destructive. Once we understand the problem, it becomes clear that the costs are too high for us to accept that this behavior should continue. Unfortunately, there will also be times when taking sides about descriptions of the real world will be the best decision, given the circumstances. When people argue against vaccines, or that evolution is only a theory (the example on General Relativity should make it abundantly clear that some theories become even more probably than any single experimental result), or any number of ``crazy'' ideas we see everyday, we often have too much data showing that they are wrong. Under some circumstances, we are actually in the cases where one of the probabilities has gone to basically zero. And then, accepting the death of little children or the spread of unsupported world views, just for the sake of a trivial reminder that $10^{-100}$ is technically (and only technically) not zero, would make no sense at all. Public discussion is a realm of incompetence and, in those case, simply stating the competent view will mean taking sides. Sadly a lot of incompetence comes from very intelligent minds who have just not educated themselves properly on everything they should know. 

Science is hard and any field takes years to master. If you are going to speak about it, it should be a moral obligation to actually know what you are talking about and why people disagree with you. In particular, when you take sides, our human nature makes you stupid and you should remember that there is evidence that, the more intelligent you are, the more stupid taking sides will make you. It is appalling to see the double standards that come from taking sides. For example, pointing that pharmaceutical industries are out there to make money is a basically correct statement, their claims must indeed be met with suspicion. And yet, an absurd number of people just fail to see that the alternatives they defend do exactly the same, that their proponents are also making money and gaining prestige from their defenders. This blindness to the problems of any sides you choose to defend is a consequence of you choosing those sides and suffering from the fact that our reasoning is not meant to reach the truth. Its function is to defend the ideas you chose, at the cost of the truth.

The need to change the way we debate should be clear by now. Taking sides makes us stupid. The more intelligent we are, the better equipped we are to defend our ideas and, paradoxically, the more stupid our beliefs will make us. Worse yet, our current way to believe does not just make it harder to get to the best answer, as if that was not already a terrible problem. There is very good reason to think that the concept that there must be a correct answer is at the very core of the raise of extremist positions, even when everyone starts with doubts. Belief and belonging are very probably the main motor behind ideological violence. By changing what we consider to be acceptable ways of discourse, we might also contribute strongly not only to faster advances and technologies, but also to a safer world. I believe or, more exactly, I strongly 'pelieve' that, if we start paying attention to when we and others have group defining beliefs and if we start denouncing that behavior in ourselves and in others, there is a strong chance we will diminish the problems I have identified here.

Of course, the concept that we must not take sides also has consequences on how we researchers interact with our own scientific practice, as we have seen. We have been making mistakes in all fields of knowledge. While those mistakes have luckily been completely inconsequential in areas as Physics, they have been very costly in other fields of knowledge.  Our desire to take sides is problematic and it is not an exaggeration to claim it has cost many lives, since we have serious problems on the published results of areas such as Medicine. In the Humanities, the  constant struggle between ideologies has certainly prevented the much needed advance of our knowledge. And, indeed, it actually means that it is close to impossible to figure out who might be right at any specific issue. Taking sides and defending ideas has rendered the opinion of researchers completely useless and it is hard to find neutral voices that could provide some amount of trust to their analysis. Incompetence can always be detected, of course. But when arguments become more solid and sophisticated, the fact that there is no logical and mathematical foundations we can use to check them in a way that is independent of our severely biased human argumentation actually means that it is not possible to even guess who is right. For anyone interested in the very important issues of the area, as we all should be, this is a major disaster. Work that simply describes situations is probably less prone to the problem, although the biases of the researchers might still compromise even those.

Accepting that we have no choice but to live with uncertainty might be a first step. Ideas can only be ruled out when this is a necessity for a practical decision to be made. Practical considerations do, of course, include our limited brain power and the cases where probability is so near zero that we really have nothing to worry about when discarding some ideas. But, somewhere in our descriptions of the area those ideas must be kept, except for the common cases where probabilities are absurdly small. Even then, we do have studies in the history of scientific thought where all old discarded ideas should survive, just in case we need them further along the path. Theories can indeed be so improbable that saying some of them are just plain wrong is a very minor, completely inconsequential abuse of language. It is the epistemological equivalent of treating distant objects as points in Physics. We know it is an approximation but we also know it is needed, the benefits far surpassing any possible precision problem. 

Another significant thing to remember is how the simple wish to obtain the one true answer has led the opinions of the agents to absurd extremes. The simulations are clearly not a representation of the whole process of human thinking. But the strength of the results does suggest that the difference between wishers, who want to find one true idea, and mixers, who accept that they might need to live with a mixture of ideas coming from different theories, might be a fundamental problem. Changing the model assumptions is still needed in a future work to check for the robustness of these results, of course. But the behavior we have observed in the models in this paper already poses significant evidence that this distinction is probably a very important one. That means that it is not only taking sides we should avoid. We must not even want to take sides.
As such, when you add the psychological already observed effects to the quite probable problems arising from social interactions suggested by the simulations, we must learn that we should never use techniques to choose between ideas. That is  also the case when the debate is purely scientific. Our statistical techniques must be developed and chosen to deal with the problem that arise from the fact that we are humans and quite fallible. As such, ranking ideas is not a problem, while rejecting them can make both readers and scientists victims to our own humanity.

It is also interesting to notice that some of the ideas that gave birth to the most damaging forms of Relativism we have observed are actually correct. They do not have the consequences people thought they have, however. But we do need to allow for the development of every type of theory and ideas. The crucial difference is that this actual need does not mean that all ideas are equal, not at all. Most will have probabilities associated to them that are so small that they will only survive as failed attempts at the theory-generating game. A few will have better luck, when they describe parts of the world in a reasonable way. Still, the fact that we do need to test as many theories as possible means we should indeed have theoretical scientists in every field of knowledge. We need people who only propose ideas and evaluate their predictions, with no attempt at all at checking their validity. And we need people who specialize in only checking those predictions also, people who has much less to lose when ideas turn out to be wrong. If those who run the experiments are different people, their observations and conclusions become far more trustworthy. Science must come to terms with our human shortcomings, with our desire to defend our own ideas and groups. And that means adopting new techniques to minimize those problems. Each and every change that diminishes the influence each scientist has on the acceptance of the ideas she defends, the better it will be for how much we can trust our whole enterprise.

We used to think we lived at the center of the Universe. We used to think we were not related to the animals. We used to think that we were governed by our reason and that our emotions did not have a significant impact on us. It is now time to realize that our reasoning is also deeply flawed. It does not work the way we always thought it did and we can not trust our own arguments. We can only learn to use tools to correct for the severe problems in our very fallible reasoning. Those tools have brought us this far and they are what make Science deserve some trust, while all other alternatives deserve none. Without those tools, it seems that, every time we try to reason, we lie to others and we lie to ourselves. We are that incompetent when we do not correct our biases.
 For now, I leave the reader with the commandment of the title: {\it ``Thou shalt not take sides''}.

\section*{Acknowledgments}
	The author would like to thank the Funda\c{c}\~ao de Amparo \`a Pesquisa do Estado de S\~ao Paulo (FAPESP) for partial support to this research under grant 2014/00551-0.


\begin{thebibliography}{100}
	
	\bibitem{faginetal95a}
	Fagin R, Halpern JY, Moses Y, Vardi MY.
	\newblock Reasoning about Knowledge.
	\newblock MIT Press; 1995.
	
	\bibitem{shope02a}
	Shope RK.
	\newblock Conditions and Analyses of Knowing.
	\newblock In: Moser PK, editor. The Oxford Handbook of Epistemology. Oxford
	University Press; 2002. .
	
	\bibitem{klein14a}
	Klein P.
	\newblock Skepticism.
	\newblock In: Zalta EN, editor. The Stanford Encyclopedia of Philosophy. summer
	2014 ed.; 2014. .
	
	\bibitem{jervis76a}
	Jervis R.
	\newblock Perception and Misperception in International Politics.
	\newblock Princeton University Press; 1976.
	
	\bibitem{mercier11a}
	Mercier H.
	\newblock Reasoning Serves Argumentation in Children.
	\newblock Cognitive Development. 2011;26(3):177--191.
	
	\bibitem{merciersperber11a}
	Mercier H, Sperber D.
	\newblock Why do humans reason? Arguments for an argumentative theory.
	\newblock Behavioral and Brain Sciences. 2011;34:57--111.
	
	\bibitem{kahanetal11}
	Kahan DM, Jenkins-Smith H, Braman D.
	\newblock Cultural cognition of scientific consensus.
	\newblock Journal of Risk Research. 2011;14:147--174.
	
	\bibitem{castellanoetal07}
	Castellano C, Fortunato S, Loreto V.
	\newblock Statistical Physics of social dynamics.
	\newblock Reviews of Modern Physics. 2009;81:591--646.
	
	\bibitem{galametal82}
	Galam S, Gefen Y, Shapir Y.
	\newblock Sociophysics: A new approach of sociological collective behavior:
	Mean-behavior description of a strike.
	\newblock J Math Sociol. 1982;9:1--13.
	
	\bibitem{galammoscovici91}
	Galam S, Moscovici S.
	\newblock Towards a theory of collective phenomena: Consensus and attitude
	changes in groups.
	\newblock Eur J Soc Psychol. 1991;21:49--74.
	
	\bibitem{sznajd00}
	Sznajd-Weron K, Sznajd J.
	\newblock Opinion evolution in a closed community.
	\newblock Int J Mod Phys C. 2000;11:1157.
	
	\bibitem{deffuantetal00}
	Deffuant G, Neau D, Amblard F, Weisbuch G.
	\newblock Mixing beliefs among interacting agents.
	\newblock Adv Compl Sys. 2000;3:87--98.
	
	\bibitem{martins08a}
	Martins ACR.
	\newblock Continuous Opinions and Discrete Actions in Opinion Dynamics
	Problems.
	\newblock Int J of Mod Phys C. 2008;19(4):617--624.
	
	\bibitem{bloor91a}
	Bloor D.
	\newblock Knowledge and Social Imagery.
	\newblock University of Chicago Press; 1991.
	
	\bibitem{hawkingpenrose10a}
	Hawking S, Penrose R.
	\newblock The Nature of Space and Time.
	\newblock Princeton University Press; 2010.
	
	\bibitem{kahanetal12a}
	Kahan DM, Peters E, Wittlin M, Slovic P, Ouellette LL, Braman D, et~al.
	\newblock The polarizing impact of science literacy and numeracy on perceived
	climate change risks.
	\newblock Nature Climate Change. 2012;2:732--735.
	
	\bibitem{popper59}
	Popper K.
	\newblock The Logic of Scientific Discovery.
	\newblock London, Hutchinson; 1959.
	
	\bibitem{quine53a}
	Quine WVO.
	\newblock Two dogmas of empiricism.
	\newblock Philosophical Review. 1953;60:20--46.
	
	\bibitem{duhem89a}
	Duhem P.
	\newblock La th\'eorie physique: Son Objet et sa Structure.
	\newblock Vrin, editor; 1989.
	
	\bibitem{laudan90a}
	Laudan L.
	\newblock Demystifying Underdetermination.
	\newblock In: Savage CW, editor. Scientific Theories. University of Minnesota
	Press; 1990. p. 267--297.
	
	\bibitem{earman92}
	Earman J.
	\newblock Bayes or Bust? A Critical Examination of Bayesian Confirmation
	Theory.
	\newblock MIT Press; 1992.
	
	\bibitem{maher93}
	Maher P.
	\newblock Betting on Theories.
	\newblock Cambridge, Cambridge University Press; 1993.
	
	\bibitem{jeffrey04}
	Jeffrey RC.
	\newblock Subjective Probability: The Real Thing.
	\newblock Cambridge, Cambridge University Press; 2004.
	
	\bibitem{martins06}
	Martins ACR.
	\newblock Probabilistic Biases as Bayesian Inference.
	\newblock Judgment And Decision Making. 2006;1(2):108--117.
	
	\bibitem{tenenbaumetal07}
	Tenenbaum JB, Kemp C, Shafto P.
	\newblock Theory-based Bayesian models of inductive reasoning.
	\newblock In: Feeney A, Heit E, editors. Inductive reasoning. Cambridge
	University Press.; 2007. .
	
	\bibitem{plous93}
	Plous S.
	\newblock The Psychology of Judgment and Decision Making.
	\newblock New York, McGraw-Hill; 1993.
	
	\bibitem{baron}
	Baron J.
	\newblock Thinking and Deciding.
	\newblock Cambridge University Press; 2007.
	
	\bibitem{kahneman11a}
	Kahneman D.
	\newblock Thinking, Fast and Slow.
	\newblock Farrar, Straus and Giroux; 2011.
	
	\bibitem{philipsedwards66}
	Philips LD, Edwards W.
	\newblock Conservatism in a simple probability inference task.
	\newblock Journal of Experimental Psychology. 1966;27:346--354.
	
	\bibitem{allais}
	Allais PM.
	\newblock The behavior of rational man in risky situations - A critique of the
	axioms and postulates of the American School.
	\newblock Econometrica. 1953;21:503--546.
	
	\bibitem{ellsberg61}
	Ellsberg D.
	\newblock Risk, ambiguity and the Savage axioms.
	\newblock Quart J of Economics. 1961;75:643--669.
	
	\bibitem{kahnemantversky79}
	Kahneman D, Tversky A.
	\newblock Prospect theory: An analysis of decision under risk.
	\newblock Econometrica. 1979;47:263--291.
	
	\bibitem{birnbaum2008a}
	Birnbaum MH.
	\newblock New paradoxes of risky decision making.
	\newblock Psych Rev. 2008;115(2):463--501.
	
	\bibitem{chapman69}
	Chapman LJ, Chapman JP.
	\newblock Illusory correlation as an obstacle to the use of valid
	psychodiagnostic signs.
	\newblock Journal of Abnormal psychology. 1969;74:271--280.
	
	\bibitem{hamiltonrose80}
	Hamilton DL, Rose TL.
	\newblock Illusory correlation and the maintenance of stereotypic beliefs.
	\newblock Journal of Personality and Social Psychology. 1980;39:832--845.
	
	\bibitem{gigerenzermuir11a}
	Gigerenzer G, Muir~Gray JA, editors.
	\newblock Better doctors, better patients, better decisions: Envisioning health
	care 2020.
	\newblock MIT Press.; 2011.
	
	\bibitem{watsonjohnsonlaird72a}
	Watson PC, Johnson-Laird P.
	\newblock Psychology of Reasoning: Structure and Content.
	\newblock Harvard University Press; 1972.
	
	\bibitem{tversykahneman83a}
	Tversky A, Kahneman D.
	\newblock Extension versus intuituive reasoning: The conjuction fallacy in
	probability judgement.
	\newblock Psych Rev. 1983;90:293--315.
	
	\bibitem{tverskykahneman81}
	Tversky A, Kahneman D.
	\newblock The framing of decisions and psychology of choice.
	\newblock Science. 1981;211:453--458.
	
	\bibitem{oskamp65a}
	Oskamp S.
	\newblock Overconfidence in case-study judgments.
	\newblock Journal of Consulting Psychology. 1965;29(3):261--265.
	
	\bibitem{lichtensteinfischhoff77}
	Lichtenstein D, Fischhoff B.
	\newblock Do those who know more also know more about how much they know? The
	Calibration of Probability Judgments.
	\newblock Organizational Behavior and Human Performance. 1977;3:552--564.
	
	\bibitem{dunningetal03a}
	Dunning D, Johnson K, Ehrlinger J, Kruger J.
	\newblock Why People Fail to Recognize Their Own Incompetence.
	\newblock Current Directions in Psychological Science. 2003;12:83--87.
	
	\bibitem{halletal07a}
	Hall CC, Ariss L, Todorov A.
	\newblock The illusion of knowledge: When more information reduces accuracy and
	increases confidence.
	\newblock Organizational Behavior and Human Decision Processes.
	2007;103:277--290.
	
	\bibitem{surowiecki05a}
	Surowiecki J.
	\newblock The Wisdom of Crowds.
	\newblock Anchor Books.; 2005.
	
	\bibitem{janis72a}
	Janis IL.
	\newblock Victims of Groupthink: A psychological study of foreign-policy
	decisions and fiascoes.
	\newblock Houghton Mifflin Company; 1972.
	
	\bibitem{kerretal96a}
	Kerr NL, MacCoun RJ, Kramer GP.
	\newblock Bias in judgment: Comparing individuals and groups.
	\newblock Psychological Review. 1996;103(4):687--719.
	
	\bibitem{asch55a}
	Asch S.
	\newblock Opinions and Social Pressure.
	\newblock Scientific American. 1955;November:31--35.
	
	\bibitem{asch56a}
	Asch SE.
	\newblock Studies of independence and conformity: A minority of one against a
	unanimous majority.
	\newblock Psychological Monographs. 1956;70(416):70.
	
	\bibitem{simon56}
	Simon HA.
	\newblock Rational choice and the structure of environments.
	\newblock Psych Rev. 1956;63:129--138.
	
	\bibitem{gigerenzertodd00a}
	Gigerenzer G, Todd PM, Group TAR.
	\newblock Simple Heuristics That Make Us Smart.
	\newblock Oxford University Press; 2000.
	
	\bibitem{tverskykahneman73a}
	Tversky A, Kahneman D.
	\newblock Availability: A heuristic for judging frequency and probability.
	\newblock Cognitive Psychology. 1973;5(2):207--232.
	
	\bibitem{gigerenzergoldstein96}
	Gigerenzer G, Goldstein DG.
	\newblock Reasoning the fast and frugal way: Models of bounded rationality.
	\newblock Psych Rev. 1996;103:650--669.
	
	\bibitem{johnsonlairdetal72a}
	Johnson-Laird PN, Legrenzi P, Legrenzi MS.
	\newblock Reasoning and a sense of reality.
	\newblock British Journal of Psychology. 1972;6(3):395--400.
	
	\bibitem{nickerson98a}
	Nickerson RS.
	\newblock Confirmation Bias: A Ubiquitous Phenomenon in Many Guises.
	\newblock Review of General Psychology. 1998;2(2):175--220.
	
	\bibitem{mercier11b}
	Mercier H.
	\newblock On the Universality of Argumentative Reasoning.
	\newblock Journal of Cognition and Culture. 2011;11:85--113.
	
	\bibitem{goebbertetal12a}
	Goebbert K, Jenkins-Smith HC, Klockow K, Nowlin MC, Silva CL.
	\newblock Weather, Climate, and Worldviews: The Sources and Consequences of
	Public Perceptions of Changes in Local Weather Patterns.
	\newblock American Meteorological Society. 2012;4(2):132--144.
	
	\bibitem{slovicetal91a}
	Slovic P, Flynn JH, Layman M.
	\newblock Perceived Risk, Trust, and the Politics of Nuclear Waste.
	\newblock Science. 1991;254(5038):1603--1607.
	
	\bibitem{kahan13a}
	Kahan DM.
	\newblock Ideology, Motivated Reasoning, and Cognitive Reflection.
	\newblock Judgment and Decision Making. 2013;8:407--424.
	
	\bibitem{kahan10a}
	Kahan D.
	\newblock Fixing the ccommunication failure.
	\newblock Nature. 2010;463:296--297.
	
	\bibitem{gettier63}
	Gettier E.
	\newblock Is Justified True Belief Knowledge?
	\newblock Analysis. 1963;23:121--123.
	
	\bibitem{hume}
	Hume D.
	\newblock An Inquiry Concerning Human Understanding.
	\newblock Oxford, Claredon Press; 1777.
	
	\bibitem{goodman46}
	Goodman N.
	\newblock Fact, Fiction and Forecast.
	\newblock Indianapolis, Bobbs-Merrill; 1946.
	
	\bibitem{ramsey31a}
	Ramsey F.
	\newblock Truth and probability.
	\newblock In: Foundations of Mathematics and other Logical Essays. K. Paul,
	Trench, Trubner and Co.; 1931. Reprinted in H.E. Kyburg and H.E. Smokler
	(eds.) (1980), Studies in Subjective Probability. New York: Robert Krieger.
	
	\bibitem{jaynes03}
	Jaynes ET.
	\newblock Probability Theory: The Logic of Science.
	\newblock Cambridge, Cambridge University Press; 2003.
	
	\bibitem{caticha04a}
	Caticha A.
	\newblock Relative Entropy and Inductive Inference.
	\newblock In: Erickson G, Zhai Y, editors. Bayesian Inference and Maximum
	Entropy Methods in Science and Engineering. No.~75 in AIP Conf. Proc. 707;
	2004. ArXiv.org/abs/physics/0311093.
	
	\bibitem{solomonoff64a}
	Solomonoff RJ.
	\newblock A formal theory of inductive inference. Part I.
	\newblock Information and Control. 1964;7(1):1--22.
	
	\bibitem{thaler80a}
	Thaler R.
	\newblock Toward a positive theory of consumer choice.
	\newblock Journal of Economic Behavior and Organization. 1980;1:39--60.
	
	\bibitem{bafumiherron10a}
	Bafumi J, Herron MC.
	\newblock Leapfrog representation and extremism: A study of American voters and
	their members in Congress.
	\newblock American Political Science Review. 2010;104(3):519--542.
	
	\bibitem{tileaga06a}
	Tileaga C.
	\newblock Representing the 'Other': A discurive analysis of prejudice and moral
	exclusion in talk about Romanies.
	\newblock Journal of Community \& Applied Social Psychology. 2006;16:19--41.
	
	\bibitem{galam05b}
	Galam S.
	\newblock Local dynamics vs. social mechanisms: A unifying frame.
	\newblock Europhysics Letters. 2005;70(6):705--711.
	
	\bibitem{bennaimetal03}
	Ben-Naim E, Krapivsky PL, Vazqueza F, Redner S.
	\newblock Unity and discord in opinion dynamics.
	\newblock Physica A. 2003;330:99--106.
	
	\bibitem{nizamanietal14a}
	Nizamani S, Memon N, Galam S.
	\newblock From public outrage to the burst of public violence: An epidemic-
	like model.
	\newblock Physic. 2014;416:620--630.
	
	\bibitem{galamjacobs07}
	Galam S, Jacobs F.
	\newblock The role of inflexible minorities in the breaking of democratic
	opinion dynamics.
	\newblock Physica A. 2007;381:366--376.
	
	\bibitem{hegselmannkrause02}
	Hegselmann R, Krause U.
	\newblock Opinion Dynamics and Bounded Confidence Models, Analysis and
	Simulation.
	\newblock Journal of Artificial Societies and Social Simulations. 2002;5(3):3.
	
	\bibitem{deffuantetal02a}
	Deffuant G, Amblard F, Weisbuch G, Faure T.
	\newblock How can extremism prevail? A study based on the relative agreement
	interaction model.
	\newblock JASSS-The Journal Of Artificial Societies And Social Simulation.
	2002;5(4):1.
	
	\bibitem{weisbuchetal05}
	Weisbuch G, Deffuant G, Amblard F.
	\newblock Persuasion Dynamics.
	\newblock Physica A. 2005;353:555--575.
	
	\bibitem{amblarddeffuant04}
	Amblard F, Deffuant G.
	\newblock The role of network topology on extremism propagation with the
	relative agreement opinion dynamics.
	\newblock Physica A. 2004;343:725--738.
	
	\bibitem{franksetal08a}
	Franks DW, Noble J, Kaufmann P, Stagl S.
	\newblock Extremism Propagation in Social Networks with Hubs.
	\newblock Adaptive Behavior. 2008;16(4):264--274.
	
	\bibitem{deffuant06}
	Deffuant G.
	\newblock Comparing Extremism Propagation Patterns in Continuous Opinion
	Models.
	\newblock JASSS-The Journal Of Artificial Societies And Social Simulation.
	2006;9(3):8.
	
	\bibitem{mckeownsheehy06a}
	Mckeown G, Sheehy N.
	\newblock Mass Media and Polarization Processes In the Bounded Confidence Model
	of Opinion Dynamics.
	\newblock JASSS. 2006;9(1):11.
	
	\bibitem{boccara10a}
	Boccara N.
	\newblock Evolution of Extremist Opinions in a Population of Interacting
	Agents.
	\newblock Int J Mod Phys C. 2010;21(5):617--628.
	
	\bibitem{alizadeh14a}
	Alizadeh M, Coman A, Lewis M, Cioffi-Revilla C.
	\newblock Integroup Conflict Escalations LLead to More Extremism.
	\newblock JASSS-The Journal Of Artificial Societies And Social Simulation.
	2014;14(4):4.
	
	\bibitem{gargiulomazzoni08a}
	Gargiulo F, Mazzoni A.
	\newblock Can Extremism Guarantee Pluralism?
	\newblock JASSS-The Journal Of Artificial Societies And Social Simulation.
	2008;11(4):9.
	
	\bibitem{martins12b}
	Martins ACR.
	\newblock Bayesian updating as basis for opinion dynamics models.
	\newblock AIP Conf Proc. 2012;1490:212--221.
	
	\bibitem{martins08b}
	Martins ACR.
	\newblock Mobility and Social Network Effects on Extremist Opinions.
	\newblock Phys Rev E. 2008;78:036104.
	
	\bibitem{eguiluzetal15a}
	Egu\'iluz VM, Masuda N, Fern\'andez-Gracia J.
	\newblock Bayesian Decision Making in Human Collectives with Binary Choices.
	\newblock PLOS ONE. 2015;10(4):e0121332.
	
	\bibitem{edwards68}
	Edwards W.
	\newblock Conservatism in human information processing.
	\newblock In: Kleinmuntz B, editor. Formal Representation of Human Judgment.
	John Wiley and Sons; 1968. .
	
	\bibitem{martins13c}
	Martins ACR.
	\newblock Discrete Opinion models as a limit case of the CODA model.
	\newblock Physica A. 2013;395:352--357.
	
	\bibitem{fanpedrycz15a}
	Fan K, Pedrycz W.
	\newblock Emergence and spread of extremist opinions.
	\newblock Physica A. 2015;436:87--97.
	
	\bibitem{martins08c}
	Martins ACR.
	\newblock Bayesian Updating Rules in Continuous Opinion Dynamics Models.
	\newblock Journal of Statistical Mechanics: Theory and Experiment.
	2009;2009(02):P02017.
	\newblock ArXiv:0807.4972v1.
	
	\bibitem{martinsgalam13a}
	Martins ACR, Galam S.
	\newblock The building up of individual inflexibility in opinion dynamics.
	\newblock Phys Rev E. 2013;87:042807.
	\newblock ArXiv:1208.3290.
	
	\bibitem{diaoetal14a}
	Diao SM, Liua Y, Zeng QA, Luo GX, Xiong F.
	\newblock A novel opinion dynamics model based on expanded observation ranges
	and individuals' social influences in social networks.
	\newblock Physica A. 2014;415:220--228.
	
	\bibitem{martins13b}
	Martins ACR.
	\newblock Trust in the CODA model: Opinion Dynamics and the reliability of
	other agents.
	\newblock Physics Letters A. 2013;377(37):2333--2339.
	\newblock ArXiv:1304.3518.
	
	\bibitem{sietal10a}
	Si XM, Liua Y, Xionga F, Zhang YC, Ding F, Cheng H.
	\newblock Effects of selective attention on continuous opinions and discrete
	decisions.
	\newblock Physica A. 2010;389(18):3711--3719.
	
	\bibitem{martins14a}
	Martins ACR.
	\newblock Opinion Particles: Classical Physics and Opinion Dynamics.
	\newblock Physics Letters A. 2014;in press.
	\newblock ArXiv:1307.3304.
	
	\bibitem{mitroff74a}
	Mitroff II.
	\newblock Norms and Counter-Norms in a Select Group of the Apollo Moon
	Scientists: A Case Study of the Ambivalence of Scientists.
	\newblock American Sociological Review. 1974;39(4):579--595.
	
	\bibitem{feyerabend94a}
	Feyerabend P.
	\newblock Against Method.
	\newblock Verso; 1994.
	
	\bibitem{fitelsonthomason08}
	Fitelson B, Thomason N.
	\newblock Bayesians sometimes cannot ignore even very implausible theories
	(even ones that have not yet been thought of).
	\newblock Australasian Journal of Logic. 2008;6:25--36.
	
	\bibitem{clemence47a}
	Clemence GM.
	\newblock The Relativity Effect in Planetary Motions.
	\newblock Reviews of Modern Physics. 1947;19(4):361--364.
	
	\bibitem{kahanetall15a}
	Kahan DM, Hoffman D, Evans D, Devins N, Lucci E, Cheng K.
	\newblock ``Ideology'' or ``Situation Sense''? An Experimental Investigation of
	Motivated Reasoning and Professional Judgment.
	\newblock University of Pennsylvania Law Review,. 2015;164:forthcoming.
	
	\bibitem{wigner60a}
	Wigner E.
	\newblock The Unreasonable Effectiveness of Mathematics in the Natural
	Sciences.
	\newblock Communications in Pure and Applied Mathematics. 1960;13(1).
	
	\bibitem{merton79a}
	Merton RK.
	\newblock 13: The Normative Structure of Science.
	\newblock In: Storer NW, editor. The Sociology of Science: Theoretical and
	Empirical Investigations. University of Chicago Press; 1979. p. 267--278.
	
	\bibitem{martins10a}
	Martins ACR.
	\newblock Modeling Scientific Agents for a Better Science.
	\newblock Adv Compl Sys. 2010;13:519--533.
	
	\bibitem{martins13a}
	Martins ACR.
	\newblock Modelling Epistemic Systems.
	\newblock In: Dabbaghian V, Mago VK, editors. Theories and Simulations of
	Complex Social Systems. Springer; 2013. .
	
	\bibitem{ioannidis05a}
	Ioannidis JPA.
	\newblock Contradicted and initially stronger effects in highly cited clinical
	research.
	\newblock Journal of the American Medical Association. 2005;294:218--228.
	
	\bibitem{nuzzo14a}
	Nuzzo R.
	\newblock Statistical Errors.
	\newblock Nature. 2014;506:150--152.
	
	\bibitem{altman94a}
	Altman DG.
	\newblock The scandal of poor medical research.
	\newblock British Medical Journal. 1994;308:283--284.
	
	\bibitem{gigerenzermarewski15a}
	Gigerenzer G, Marewski JN.
	\newblock Surrogate Science: The Idol of a Universal Method for Scientific
	Inference.
	\newblock Journal of Management. 2015;41(2):421--440.
	
	\bibitem{cumming08a}
	Cumming G.
	\newblock Replication and $p$ Intervals.
	\newblock Perspectives on Psychological Science. 2008;3(4):286--300.
	
	\bibitem{held10a}
	Held L.
	\newblock A nomogram for P values.
	\newblock BMC Medical Research Methodology. 2010;10:21--22.
	
	\bibitem{ioannidis05b}
	Ioannidis JPA.
	\newblock Why Most Published Research Findings Are False.
	\newblock PLOS ONE. 2005;2(8):e124.
	
	\bibitem{cumming14a}
	Cumming G.
	\newblock The New Statistics: Why and How.
	\newblock Psychological Science. 2014;25(1):7--29.
	
	\bibitem{savaleidunn15a}
	Savalei V, Dunn E.
	\newblock Is the call to abandon p-values the red herring of the replicability
	crisis?
	\newblock Frontiers in Psychology. 2015;6:245.
	
	\bibitem{trafimowmarks15a}
	Trafimow D, Marks M.
	\newblock Editorial.
	\newblock Basic and Applied Social Psychology. 2015;37:1--2.
	
	\bibitem{leekpeng15a}
	Leek JT, Peng RD.
	\newblock $P$- values are just the tip of the iceberg.
	\newblock Nature. 2015;520:612.
	
\end{thebibliography}

\end{document}